\documentclass[twocolumn]{aastex631}

\begin{document}

\title{Identifying new high-confidence polluted white dwarf candidates using Gaia XP spectra and Self-Organizing Maps}

\author[0000-0001-5797-252X]{Xabier P\'erez-Couto}
\affiliation{Universidade da Coruña (UDC), Department of Computer Science and Information Technologies,\\ Campus de Elviña s/n, 15071, A Coruña, Galiza, Spain}
\affiliation{CIGUS CITIC, Centre for Information and Communications Technologies Research,\\ Universidade da Coruña, Campus de Elviña s/n, 15071 A Coruña, Galiza, Spain}

\author[0000-0001-9296-3100]{Lara Pallas-Quintela}
\affiliation{Universidade da Coruña (UDC), Department of Computer Science and Information Technologies,\\ Campus de Elviña s/n, 15071, A Coruña, Galiza, Spain}
\affiliation{CIGUS CITIC, Centre for Information and Communications Technologies Research,\\ Universidade da Coruña, Campus de Elviña s/n, 15071 A Coruña, Galiza, Spain}

\author[0000-0002-7711-5581]{Minia Manteiga}
\affiliation{Universidade da Coruña (UDC), Department of Nautical Sciences and Marine Engineering,\\ Paseo de Ronda 51, 15011, A Coruña, Galiza, Spain}
\affiliation{CIGUS CITIC, Centre for Information and Communications Technologies Research,\\ Universidade da Coruña, Campus de Elviña s/n, 15071 A Coruña, Galiza, Spain}

\author[0000-0003-4936-9418]{Eva Villaver}
\affiliation{Instituto de Astrofísica de Canarias, 38200 La Laguna, Tenerife, Spain}
\affiliation{Universidad de La Laguna (ULL), Astrophysics Department,\\ 38206 La Laguna, Tenerife, Spain}

\author[0000-0003-4693-7555]{Carlos Dafonte}
\affiliation{Universidade da Coruña (UDC), Department of Computer Science and Information Technologies,\\ Campus de Elviña s/n, 15071, A Coruña, Galiza, Spain}
\affiliation{CIGUS CITIC, Centre for Information and Communications Technologies Research,\\ Universidade da Coruña, Campus de Elviña s/n, 15071 A Coruña, Galiza, Spain}



\received{July 26, 2024}
\accepted{October 18, 2024}

\begin{abstract}

The identification of new white dwarfs (WDs) polluted with heavy elements is important since they provide a valuable tool for inferring chemical properties of putative planetary systems accreting material {on} their surfaces. The Gaia space mission has provided us with an unprecedented amount of astrometric, photometric, and low resolution (XP) spectroscopic data for millions of newly discovered stellar sources, among them thousands of WDs. 
In order {to find WDs among this data} and to identify which ones have metals in their atmospheres, we propose a methodology based on an unsupervised artificial intelligence technique called Self-Organizing Maps (SOM). In our approach a nonlinear high-dimensional dataset is projected on a 2D grid map where similar elements fall into the same neuron.

By applying this method, we obtained a clean sample of 66,337 WDs. We performed an automatic spectral classification analysis to them, obtaining 143 bona fide polluted WD candidates not previously classified in the literature. The majority of them are cool WDs and we identify in their XP spectra several metallic lines such as Ca, Mg, Na, Li, and K. The fact that we obtain similar precision metrics than those achieved with recent supervised techniques highlights the power of our unsupervised approach to mine the Gaia archives for hidden treasures to follow-up spectroscopically with higher resolution.
\end{abstract}
\keywords{white dwarfs --- methods: data analysis --- catalogs}


\section{Introduction} \label{sec:intro}

White dwarfs (WD) are the degenerate stellar remnants of low- to intermediate- mass stars ($\leq 8 M_{\odot}$) \citep{ibenetal1997}. Due to their high density ($\sim 10^3$ kg/m$^3$), WDs have fully stratified interiors, containing a degenerate core composed of carbon and oxygen. This core is encased in a thin helium mantle, which constitutes at most about 1\% of the white dwarf's mass. Surrounding this helium layer is an even thinner but opaque hydrogen envelope, which makes up no more than approximately 0.01\% of the mass. In some cases, Carbon can also be detected in the atmosphere, diffused upwards from the nuclei by a convection zone under the Helium layer \citep{pelletieretal1986}. 

More exciting are those WDs that show heavy metal lines in their atmospheres. In cool WDs (below approximately 25,000 K), heavy elements tend to diffuse downward in the atmospheres due to gravitational settling in the presence of strong gravitational fields \citep{koester2009}. Since the diffusion timescales due to gravitational settling are much shorter than the evolutionary time scales of WDs, those metals cannot be primordial; they must have been accreted, with the accretion of rocky material from planetesimals being the most widely accepted explanation  \citep{koesteretal2006, zuckermanetal2007, farihietal2010, verasetal2024}. For this reason, the detection of those polluted WDs is nowadays an effervescent field and a valuable tool to infer the presence and physical properties of exoplanets \citep{mustilletal2018, izquierdoetal2020, maldonadoetal2020, maldonadoetal2021, kleinetal2021, trierweileretal2023, swanetal2023, xuetal2024}.

WD cooling tracks start at high temperatures ($\sim{10^5}K$) and thus they are located in the blue part of the color-magnitude diagram (CMD). The low luminosity resulting from their small radius (approximately 10,000 km) \citep{carvalhoetal2016} makes WDs very difficult to observe. In fact, as of 2018, only about 30,000 WDs had been identified.  {Fortunately, the Gaia mission has changed the rules of the game by providing 
$G_{BP} - G_{RP}$ colors from Blue ($BP$) and Red ($RP$) photometers, as well as parallaxes and full astrometry, for 1460 million sources in the Gaia Data Release 3 (Gaia DR3) \citep{gaiacollaboration2023}}. Gaia low resolution BP/RP (XP) mean spectra {are} also provided for $\sim 220$ million sources by using spectrophotometry \citep{carrascoetal2021}.

Since classifying such a large number of spectra by human visual inspection is not feasible, several works took advantage of supervised machine learning techniques to train algorithms with labeled WD spectra and, subsequently, to predict the spectral classes of the rest of unlabeled objects. In \citet{vincentetal2023} and \citet{vincentetal2024} a neural network was trained with $\sim 14,000$ Sloan Digital Sky Survey (SDSS) WDs in order to predict if a {\citet{gentilefusillo2021} (hereafter, GF+21)} source is a WD or a contaminant. {Subsequently, the sources are classified} in one of the primary spectral classes: DA (H - rich), DB (He I - rich), DO (He II - rich), DQ (C - rich), DZ (metal - rich) or DC (featureless) WDs, \citep{sionetal1983} with precisions above $90\%$ {for DA, DB, and DO types, but lower for DC, DQ, and DZ.} \citet{garciazamoraetal2023} used the Random Forest algorithm with excellent precision ($>90\%$) for primary classes and less accurate results, arguably due to the unbalance of the training set, for secondary classes where traces of other elements are expected (DAZ, DBZ, \dots).

Indeed, a key characteristic of supervised machine learning techniques is their complete reliance on the training dataset. Thus, the performance is severely affected if the training dataset is unbalanced with underrepresented classes. Furthermore, several works use training datasets built from SDSS spectra with a higher resolution ($R\approx 1800$) than that of Gaia, {and as a consequence, those artificial intelligence models can misinterpret some non atmospheric features in the XP spectra such as the observational noise.} These two reasons strongly justify exploring alternative unsupervised learning approaches. Recently, \citet{kaoetal2024}, used an unsupervised dimensionality reduction technique called Uniform Manifold Approximation and Projection (UMAP) {to classify the Gaia XP spectra of WDs with a previous $P_{WD} > 0.9$ filter in the GF+21 catalogue. The UMAP allowed them to project each vector of 110 XP coefficients in a 2D map or manifold where similar elements appears next to each other. Subsequently, they found a well defined island of similar WD candidates showing Ca and other metals in their Gaia XP spectra. As a result, their methodology allowed them to discover new 375
polluted white dwarfs candidates.} We will discuss these works in more detail later and compare their results with our own.

In this work, we use a neural network-based dimensionality reduction algorithm called Self-Organizing Maps (SOM) \citep{kohonen1982} where, given a nonlinear high-dimensional dataset, the input data is projected on a 2D grid map where similar elements fall into the same neuron. Here, the similarity is defined by a metric (e.g. Euclidean distance) so the unsupervised learning process aims to maximize the similarity between objects belonging to the same neuron at the same time it minimizes the similarity between objects within different neurons. Therefore, topology is naturally preserved: similar neurons are also grouped next to each other. Once the clustering process ends, each neuron is labeled by means of a template-matching procedure where a tag is assigned to the mean element of that neuron (the prototype) \citep{delchambreetal2023, pallasquintela2023}. 

Consequently, SOMs combine the two major utilities of unsupervised learning: dimensionality reduction and clustering, unlike other algorithms {that either perform only clustering (e.g. K-means) or only dimensionality reduction (e.g. t-SNE, UMAP).} This double characteristic demonstrated SOMs to be a great artificial intelligence tool for object classification in different fields of Astronomy and Astrophysics \citep{torresetal1998, naimetal2009, geach2012, wayandklose2012, carrascoandbrunner2014, alvarezetal2022}, as well as for outlier detection and analysis \citep{ordonezetal2010, fustesetal2013a, fustesetal2013b, dafonteetal2018}.

The manuscript is organised as follows: in \S\ref{sec:methodology} we present the methodology to filter out and classify the Gaia DR3 WD candidates with SOM. The key results of our unsupervised classification procedure are shown in \S\ref{sec:results} and the main conclusions of this work are summarized in \S\ref{sec:conclusions}.

\section{Methodology} \label{sec:methodology}

{Most of the Gaia DR3 sources} can be placed on an absolute $G$ magnitude to $G_{BP} - G_{RP}$ color diagram. In \citet{gentilefusillo2019} and \citet{gentilefusillo2021} this CMD is used to identify WDs in Gaia Early DR3 by applying color and absolute magnitude cuts that correspond to the WD region in the Hertzsprung–Russell diagram.

Since belonging to that region is a necessary but not a sufficient condition to be a WD (e.g. QSOs are also blue and faint), \citet{gentilefusillo2021} determined a probability of being a WD ($P_{WD}$) as a function of the position in the color-magnitude parameter space, by using a well-known sample of {$22,998$} spectroscopically confirmed WDs and {$7124$} contaminants identified by visual inspection in the SDSS and subsequently converted to normalized 2D Gaussians in order to obtain two continuous density maps (one for confirmed WDs and other for contaminants). Then the $P_{WD}$ for a given candidate was computed by integrating its CMD Gaussian distribution with the map resulting of taking the ratio of the confirmed WD density map to the sum of both maps.

As a result of applying this methodology to Gaia EDR3, \citet{gentilefusillo2021} computed a catalog of around $1.3$ million candidates, of which $\approx 359,000$ are classified as high-confidence WDs by imposing a $P_{WD} > 0.75$ probabilistic cut. The GF+21 sample of WD candidates not only increased the number of WDs by an order of magnitude, but it also allowed to classify such a big sample in different spectral types by means of their Gaia XP spectra.

Gaia XP spectra are provided for $\sim 220$ million sources from mean low resolution ($R \approx 70$) spectrophotometry {up-to-date}, and therefore represents a valuable tool to study the astrophysical properties of a large amount of objects. Instead of fluxes per wavelength unit, Gaia provides 110 coefficients (55 for the BP and 55 for the RP spectra) corresponding to sets of basis functions resulting from multiplying Gaussian functions to Hermite polynomials \citep{carrascoetal2021}. These coefficients uniquely define the spectra of each source, and therefore they can be used to study its spectral features \citep{weileretal2023}. 

\subsection{Initial sample}\label{initialsample}

{As a baseline, we use GF+21 sample}, that is constructed by imposing the following color-magnitude cut to isolate the WD locus in the CMD:
\begin{equation}\label{WDlocus}
    G_{abs} > 6 + 5 \cdot \left(G_{BP} - G_{RP}\right)
\end{equation}
Subsequently, a large list of quality filters over several astrometric and photometric parameters was applied until arriving at a sample of $1,280,266$ sources within the WD region with high-quality measurements (see \S2.1. in \citet{gentilefusillo2021} for more information about all of those quality filters). 

Finally, since we completely rely on the morphology to classify the spectra, we have applied several additional filters to ensure the quality of the input data: i)
\texttt{(phot\_bp\_n\_obs > 10) \& (phot\_rp\_n\_obs > 15)}, refer to the minimum number of CCD transits for BP and RP spectra, respectively, {based on the recommendations in \citet{andraeetal2023} to ensure a sufficient signal-to-noise ratio (S/N) for posterior spectral analysis. We consider that the asymmetry in the cuts for BP and RP minimum number of observations are suitable for WDs since the weakest metallic lines we aim to find, such as K I or Li I, are in the RP wavelength range, and therefore a better signal-to-noise ratio is required to identify them in comparison with the BP lines (Ca II, Na I) that have more S/N.}; ii) \texttt{visibility\_periods\_used > 10}, being each visibility period a group of observations separated from the next one by at least 4 days, keeping in this way only those sources that have been astrometrically ``well-observed'' \citep{lindegrenetal2018}; iii) \texttt{|phot\_bp\_rp\_excess\_factor\_corrected| < 5 x sigma\_excess\_factor} which, as explained in \citet{rielloetal2021}, makes sure that the photometry of $G_{BP}$, $G_{RP}$, and $G$ is consistent, avoiding any possible contamination from external sources in the same field-of-view.

Then, we retrieved the Gaia XP spectra of this sample by using the DataLink Gaia tool (\url{https://www.cosmos.esa.int/web/gaia-users/archive/datalink-products}) through the \texttt{astroquery} Python package \citep{ginsburgetal2019}, downloading  XP coefficients for $104,844 $ sources.

\subsection{Kohonen's Self-Organizing Maps}

{In contrast to other works where the GF+21 catalog is used, here we did not appeal to the $P_{WD}$ parameter to filter out contaminants. While we agree that this probabilistic cut discards the majority of contaminants from the initial sample, it relies only on the $G$, $G_{BP}$, and $G_{RP}$ magnitudes. Since the XP coefficients encode much more information about the astrophysical properties of the Gaia sources, such as absorption and emission lines (key features to identify contaminants such as QSOs, \citealt{deangelietal2023}), they are more valuable to obtain a more conservative sample of bona fide WDs. To tackle this issue, we applied our own SOM approach to cluster WDs and contaminants in different neurons.}

The learning process of SOM, as explained in the original implementation of \citet{kohonen1982}, starts with a random initialization of the weights $w_{m,n}$ for each neuron $z_{m,n}$, being $m\times n$ the predefined dimension of the map. Then, for each input element $x_i$, an initial iteration looks for the winner neuron (best matching unit, or BMU) by minimizing the Euclidean distance between $x_i$ and each $w_{m,n}$. Subsequently, an iterative process updates the weights by means of a neighborhood function that preserves the topology with closer neurons. The learning process stops when the weights do not change significantly between iterations.

In pursuit of applying the SOM algorithm to the selected spectra, we chose the \texttt{MiniSom}\footnote{\url{https://github.com/JustGlowing/minisom/}} library in Python due to its simplicity and hyperparameter flexibility \citep{vettigli2018}. After some hyperparameter tuning, we {selected} the Euclidean as the activation distance, the Gaussian as the neighborhood function, and an initial learning rate of 0.7. Another important parameter is $\sigma$, which gives the initial spread of the neighborhood function so that a greater $\sigma$ will consider farther neurons as neighbours while a smaller $\sigma$ will only look at the closer ones. Finally, it should be noticed that such as any algorithm that depends on a distance calculation, it is very sensitive to the scale of the features (here, the XP coefficients), so to ensure a good performance we normalize each input vector of XP coefficients by dividing it by its $\mathcal{L}_2$ norm before the learning process.

\subsection{Filtering out of contaminants}\label{filtering}

\begin{figure}
    \centering
    \includegraphics[width=0.49\textwidth]{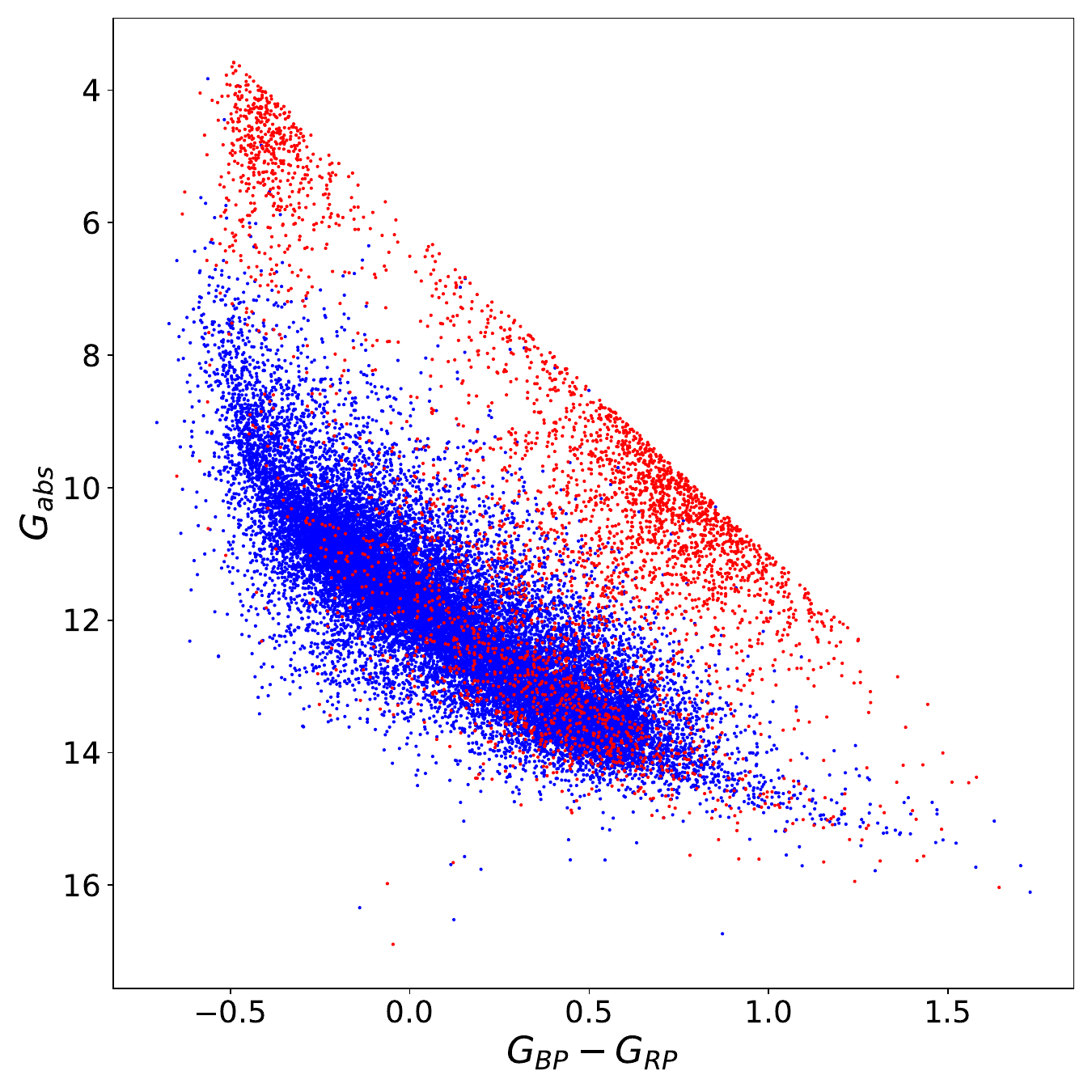}
    \caption{CMD of Gaia-SDSS spectroscopically confirmed white dwarfs (blue) and contaminants (red).}
    \label{fig:gaia-sdss_CMD}
\end{figure}

In order to clean our initial sample from non-WD objects that could share the locus {defined in Equation} \eqref{WDlocus} with true WDs, we use the same Gaia-SDSS spectroscopic sample published in GF+21 {and available online \footnote{\url{https://warwick.ac.uk/fac/sci/physics/research/astro/research/catalogues/gaiaedr3_wd_sdssspec.fits.gz}}}. This sample contains $42,007 $ SDSS DR16 spectra of $32,321$ sources. Here we only use those that pass the quality filters mentioned in \S\ref{initialsample} and with available XP spectra. {As a result, we obtained} a final sample of $10,835$ sources, divided in $10,141$ confirmed WDs and $694$ contaminants labeled with any of the following tags: CV, DB\_MS, DA\_MS, DA\_MS:, DC\_MS, STAR, QSO, Unreli, UNKN, GALAXY, DA\_DQ. In Figure \ref{fig:gaia-sdss_CMD} the Gaia-SDSS spectroscopically confirmed WDs (blue) and contaminants (red) are shown in a CMD.

To obtain reliable statistics we need to use synthetic data, since the sample size of contaminants is much smaller than that of confirmed WDs, with a poor ratio of 0.06. For this purpose, we use the Synthetic Minority Oversampling Technique (SMOTE) through the \texttt{imbalanced-learn}\footnote{\url{https://imbalanced-learn.org}} Python library \citep{lemaitre2017} over the normalized XP coefficients, with a resampling ratio of 0.20 for the minority class ($10,141$ confirmed WDs versus 2028 contaminants). The random seed for SMOTE as well as for the random initialization of the following SOMs was set to 1. With this resampled dataset we proceeded to train a SOM by choosing a small map size of $5\times5$ in order to have enough stars within each neuron. The number of iterations was $5\times 10^4$, with $\sigma = 1.5$, and a learning rate of 0.5. 

\begin{figure}
    \centering
    \includegraphics[width=0.495\textwidth]{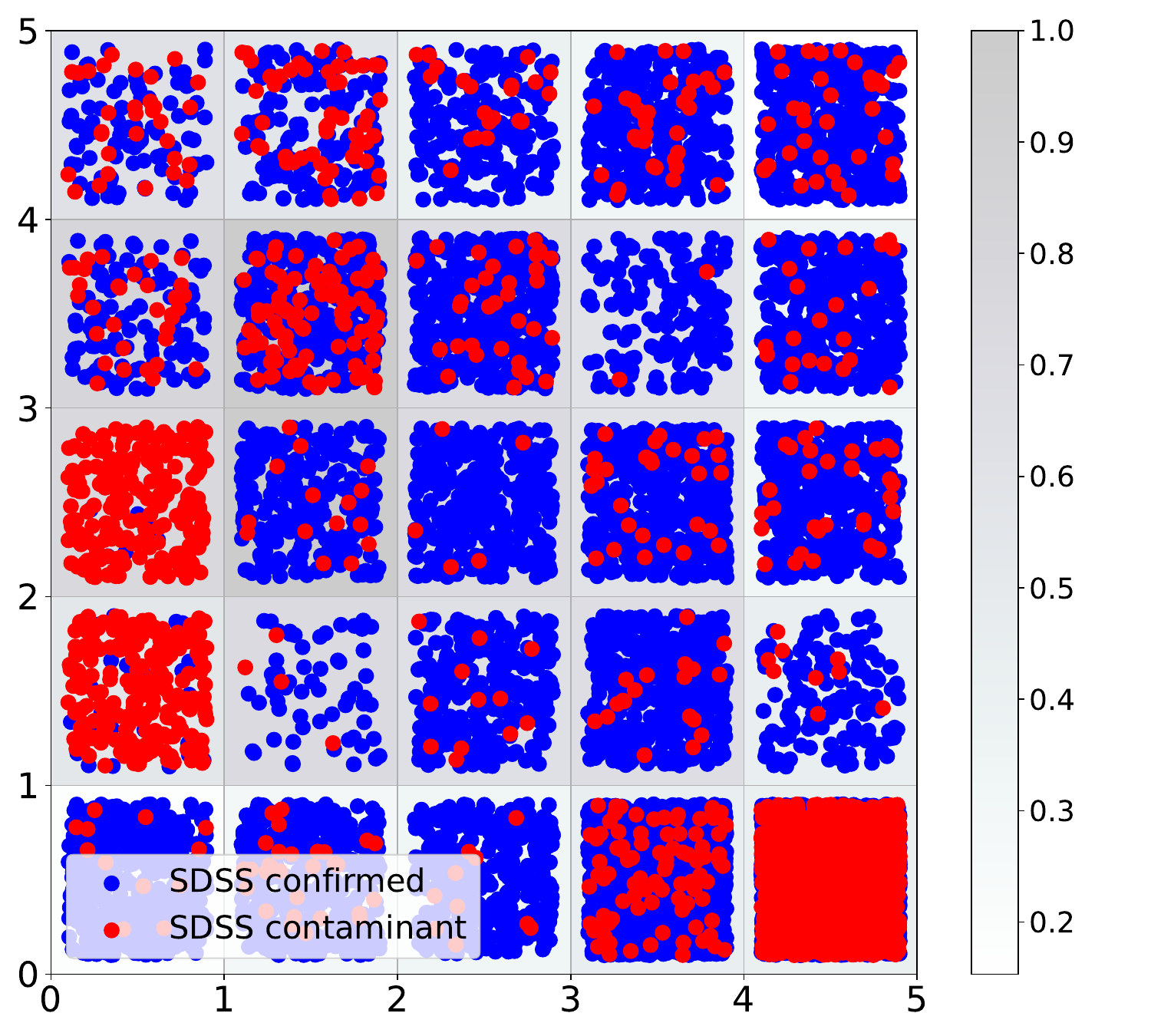}
    \caption{Self-Organized map of the Gaia-SDSS spectroscopic sample after $5\times 10^4$ iterations. SDSS confirmed WDs are shown in blue and SDSS and synthetic contaminants in red. The {side bar} shows the Intra-Neuron Distance (IND) or how similar are the objects within the same neuron.}
    \label{fig:gaiasdsss_som}
    
\end{figure}

The Self-Organized Map obtained is shown in Figure \ref{fig:gaiasdsss_som}, where contaminants are depicted in red colors. As can be seen in the figure, there are neurons with different levels of contamination. While in the neuron $z_{0,4}$ the confirmed WDs are indistinguishable from the contaminants, $z_{3,3}$ just has a few of them. Thus, {if we define the purity of a neuron as the ratio of the number of spectroscopically confirmed WDs to the total number of sources within that neuron, out of 25 neurons, 18 have a purity level greater than 90\%.} 

A total of 7338 sources ($\sim 60\%$) fell into those neurons, of which the expected $<10\%$ of contaminants correspond to 150 SDSS labeled sources completed with 269 unlabeled synthetic contaminants. From the labeled SDSS sources, only 18 are confirmed stars (STAR), 79 correspond to SDSS sources with unreliable spectral class, 1 is tagged as UNKN and the rest of them are CV (19 sources) and WD-main sequence binaries (33 sources). Therefore, no contaminants such as QSOs or galaxies are expected by using our method.

Once the map has been pre-trained (meaning that the final weight of each neuron has been computed with the XP coefficients of the Gaia-SDSS sample) we pass the initial sample defined in subsection \ref{initialsample} through the map, so that each vector of XP coefficients falls within the neuron to which the weight is closest. As a result, $66,337$ sources ($\sim 63\%$) fell into our pure neurons.

\subsection{Spectral classification}

\begin{figure}
    \centering
    \includegraphics[width=0.495\textwidth]{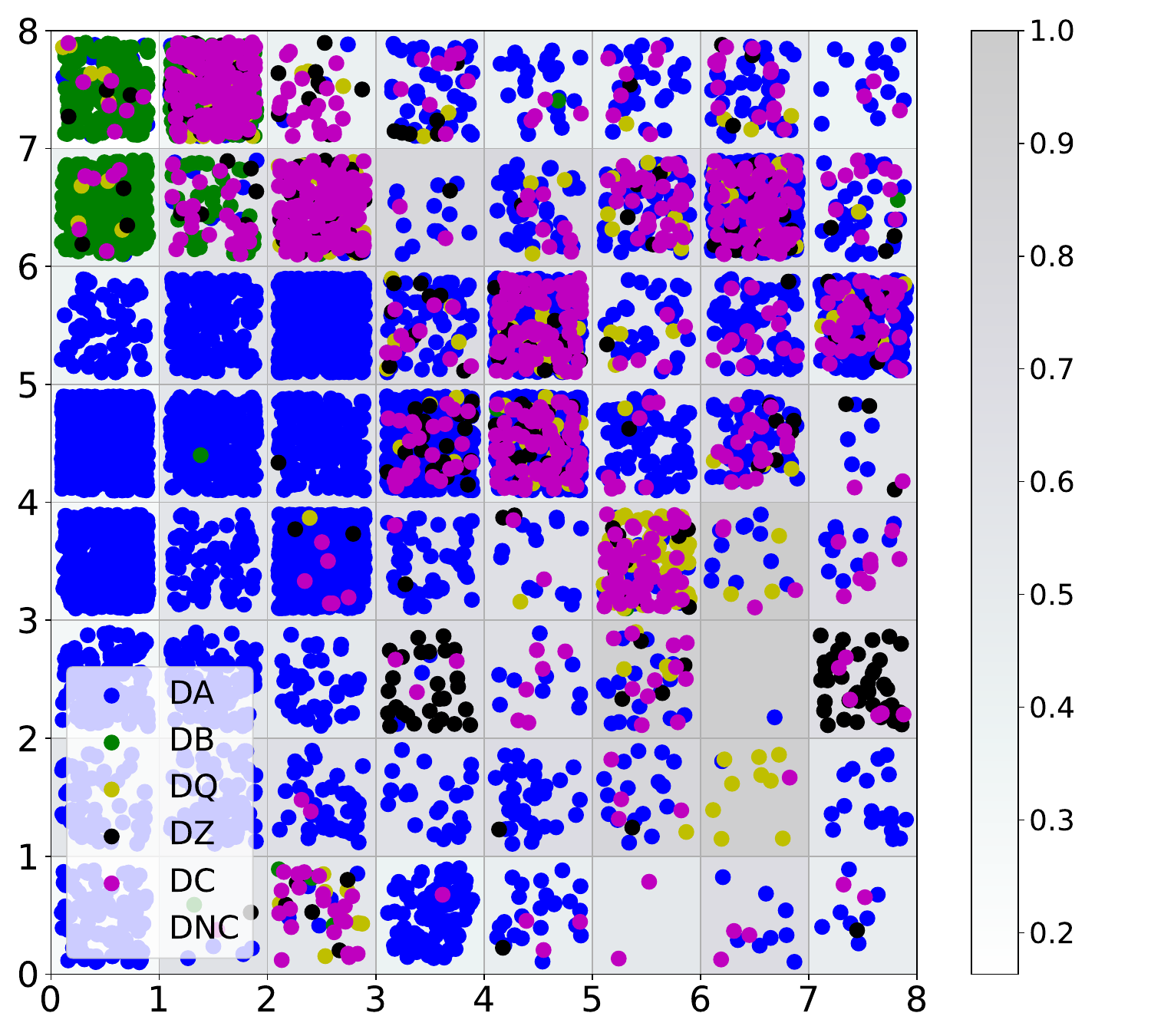}
    \caption{SOM map with our sample of $66,337$ clean sources. WDs with known spectral type in MWDD catalog appear in different colors, and candidates are invisible to enhance visualization.}
    \label{fig:som_candidates}
\end{figure}

In order to classify the Gaia WD spectra, we have passed the $66,337$ clean sources {selected as WDs} through a $8\times 8$ SOM with the same hyperparameters as in \S\ref{filtering} except that we use $\sigma =1$ here, so that the algorithm discriminates more and thus increases precision.With the aim to give labels to each neuron, we cross-matched this sample with the Montreal White Dwarf Database (MWDD)\footnote{\url{https://www.montrealwhitedwarfdatabase.org/}} \citep{dufouretal2016} and separate them in the {six primary spectral classes: DA, DB, DO, DQ, DZ and DC, referring to the main chemical abundances present in their atmospheres. The resulting map is shown in Figure \ref{fig:som_candidates}, where only the 9273 WDs with confirmed classification in the MWDD are colored,} {being 7088 DAs, 933 DCs, 634 DBs, 340 DZs, and 278 DQs. There are not DOs after the SOM filter, so from now on we will not deal with them. The remaining $57,064$ DNC (short for `Not Classified') candidates are invisible for the sake of visualization.}

Subsequently, we used the most frequent class within each neuron as its representative class, {if and only if} the ratio of that class with respect to the others is of 0.5 or higher. {We have chosen 0.5 because we aim to use the most frequent class as the representative label of that neuron but, at the same time, there are neurons where the ratio of the classes is too spread out to state with certainty that all the stars in that neuron belong to the most frequent class. Therefore, setting a minimum threshold is highly recommended.} Otherwise, the neuron is classified as Outlier since it contains too mixed classes and no high-precision classification would be reliable. By applying this filter, 60 neurons are considered useful for classification, accounting for 61,817 WDs, and only four of them ($z_{1,6}$, $z_{1,7}$, $z_{5,2}$, and $z_{5,3}$) are labeled as Outlier.

\section{Results and discussion}\label{sec:results}

\subsection{WD candidates}\label{subsec:WDcandidates}
Regarding the filter of contaminants applied in this work, from the $66,337$ clean sources, 1804 have a $P_{WD} < 0.9$ while 791 sources have $P_{WD} < 0.75$ in \citet{gentilefusillo2021}, showing that the use of {Gaia XP coefficients} instead of Gaia CMD has allowed us to reveal new candidates. {Furthermore,  84\% of the contaminants are sources that are considered in the GF+21 catalog as high-confidence WD candidates (they have $P_{WD}> 0.75$). We certainly have imposed a quite conservative filter in order to get the purest sample. Therefore, in exchange for precision, it is expected that we are considering quite a few true WDs as contaminants. However, we also find that a very important fraction of the high confidence WD candidates in GF+21 have XP spectra morphologically similar to contaminants such as stars and QSOs. Thus, care must be taken.}

{One of the main limitations of the SOM is its recall, since the objective of this work is to obtain a the most pure sample, and therefore conservative filters are prioritized}. As a consequence, in the sample of 38,507 sources classified as contaminants in \S\ref{filtering}, there are 14,376 ($37\%$) sources in the MWDD and 5844 ($15\%$) of them have a confirmed spectral type. Taking into account the total number of WDs from each MWDD spectral type in our initial sample, we can roughly estimate that we are missing $38\%$ of DAs, $47\%$ of DBs, $42\%$ of DCs, $19\%$ of DZs, $25\%$ of DQs, and $100\%$ of DOs by applying our SOM filter. Fortunately, the DZs that are the spectral type we are most interested in hunting are also the less affected ones.

{We applied the Equation 2 of \citet{golovinetal2024} to the sample of sources within 50pc to the Sun (see Equation 1 in the same paper) selected by the SOM filter as WDs (743 sources), in order to see if they may be contaminants due to spurious astrometric solutions. All those sources passed correctly the filter. The same was applied for the sources classified as contaminants by the SOM within the same distance (1405 sources), finding that three of them are also considered contaminants in \citet{golovinetal2024}.}

\begin{figure}
    \centering
    \includegraphics[width=0.49\textwidth]{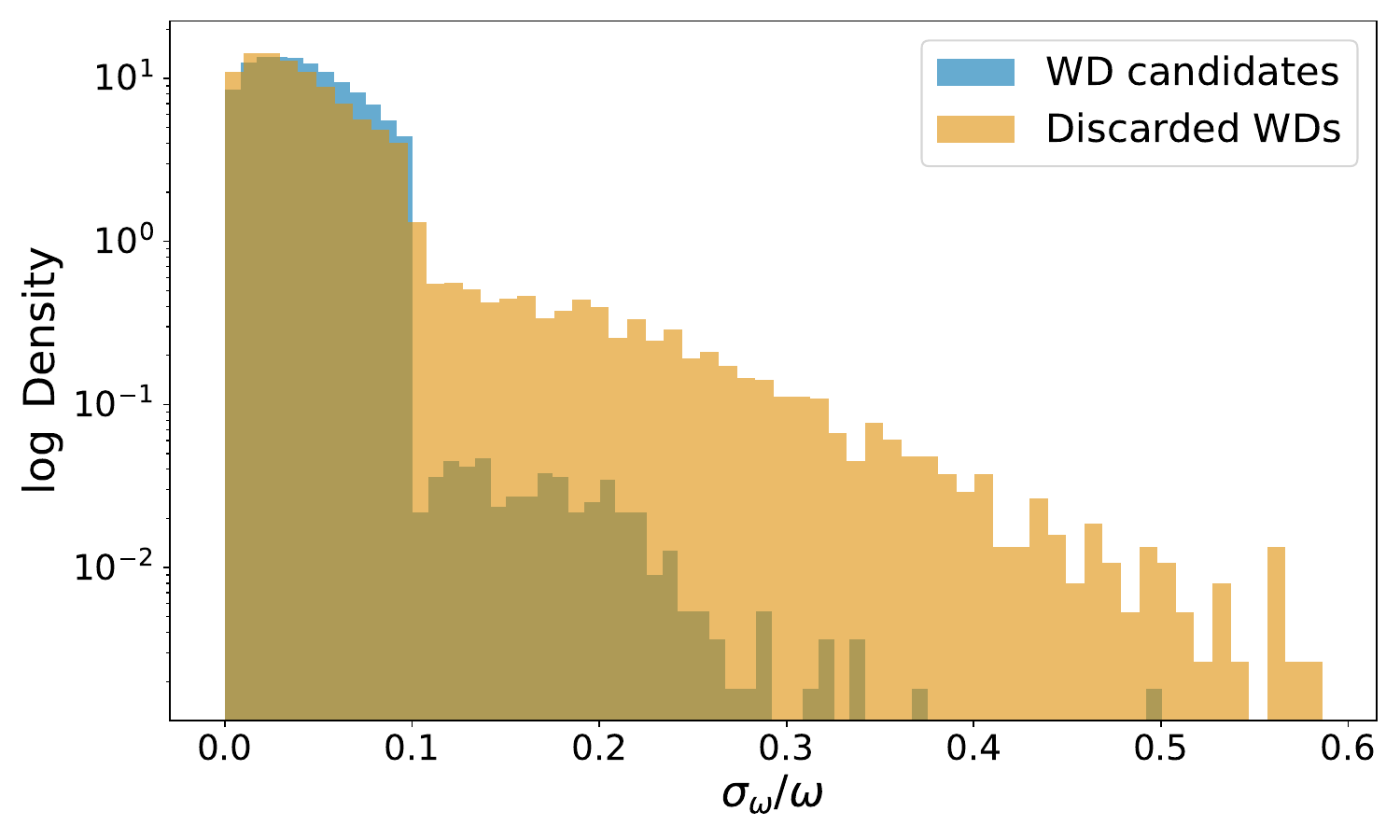}
    \caption{Parallax relative error distribution (density in log scale) of the $66,337$ WD candidates and of the $38,507$ sources classified as possible contaminants.}
    \label{fig:parallas_error_distribution}
\end{figure}

We have also studied the distribution of parallax relative errors in both the $66,337$ WD candidates and in the sample of $38,507$ sources classified as possible contaminants. 
We found that while only 291 of the WD candidates ($\sim 0.7\%$) have a parallax relative error $\left(\sigma_{\omega}/\omega\right)$ $>10\%$,  this number increases significantly to 2949 for the contaminants ($\sim 8\%$ of them). Thus, part of the contamination is most likely due to the fact that poor parallaxes introduce errors in $G_{abs}$, and as a consequence in the position of the source in the HR diagram. Notwithstanding that, and as can be seen in Figure \ref{fig:parallas_error_distribution}, most of the sources classified as contaminants show good parallaxes as well. 

{This could suggest that most of the sources classified as contaminants are actually true WDs. However, several types of contaminants, such as unresolved binaries, can still appear in the WD locus despite being correctly identified as contaminants. As a result, the parallax error cut may not be sufficient to effectively eliminate contamination in large samples of WD candidates. The data analyzed in this study is still insufficient to draw definitive conclusions.}

\subsection{Spectral classification}\label{subsec:spectralclassification}

\begin{figure}
    \centering
    \includegraphics[width=0.495\textwidth]{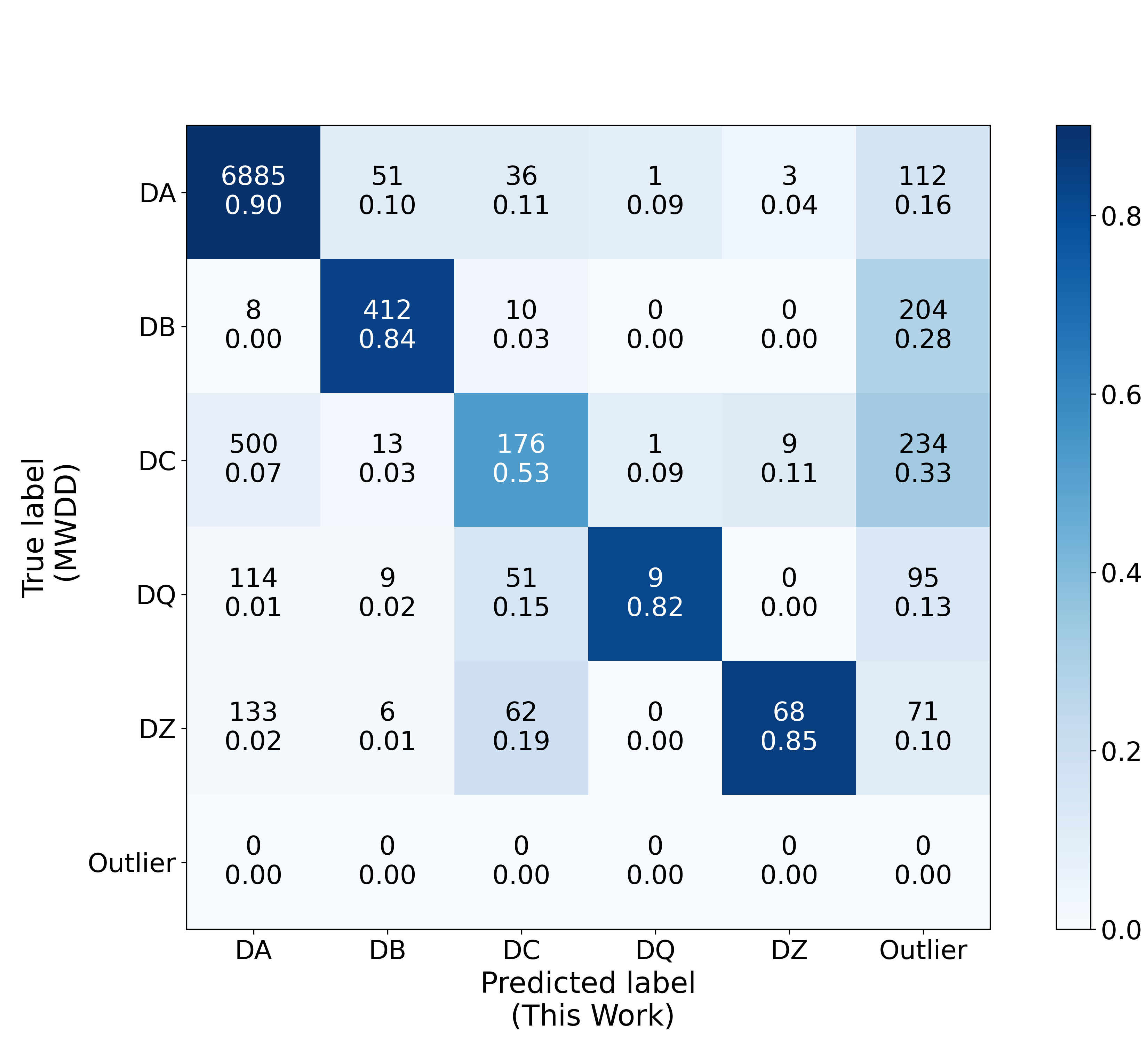}
    \caption{Confusion matrix of the SOM classification for WD primary spectral types with respect to MWDD classification.}
    \label{fig:som_confusionmatrix}
\end{figure}

Once we have assigned a label to each neuron (and therefore to each WD falling in that neuron), we can compare the predicted class with the true class given by the MWDD by using a {confusion matrix $C$ such the one shown in Figure \ref{fig:som_confusionmatrix}. The number in each cell $C_{i,j}$ shows the number of WDs with a true label $i$ and a predicted class $j$. Immediately below is shown the same quantity, but normalized over the predicted labels (columns) so the precision of the classification for each class appears in the diagonal. Thus, the confusion matrix would be diagonal for an ideal classification.} Our confusion matrix shows excellent precision for DA and DZ classes ($\geq 85\%$), very good precision for DB and DQ classes ($\geq 80\%$), and poor metrics for DC class that is mainly confused with {DAs and Outliers}. In addition, Outlier neurons are mainly populated with a mixture of all classes but mainly with DCs and DBs. Recall is excellent for DAs but poor for the rest of the classes.
In Table \ref{tab:som_metrics} the precision and recall metrics obtained for each class are summarized. 
\begin{table}
\caption{\label{tab:som_metrics} {Precision and recall metrics for each primary spectral class.}}
\begin{center}
\begin{footnotesize}
\begin{tabular}{llll}

{Class}  & {Precision} & {Recall}   \\ \hline
DA & 0.90      & 0.97    \\
DB & 0.84      & 0.65     \\
DC & 0.53      & 0.19     \\
DQ & 0.82      & 0.03     \\
DZ & 0.85      & 0.20     \\ \hline
\end{tabular}
\end{footnotesize}
\end{center}
\end{table}

\begin{figure}
\centering
    \includegraphics[width=0.48\textwidth]{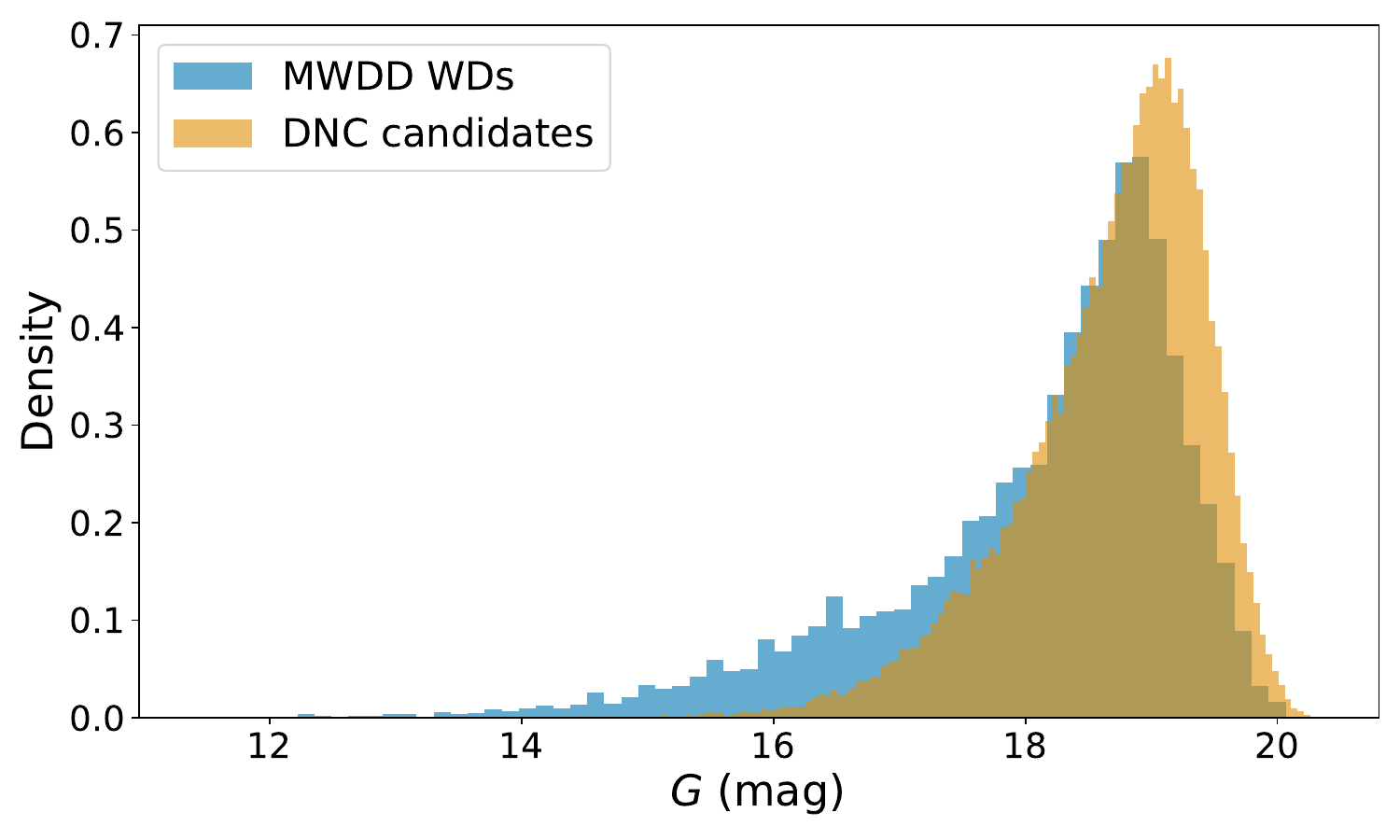}
    \caption{Density histograms of $G$-band magnitude for WDs with confirmed spectral classification and WD candidates. No differences between spectral classes were observed.}
    \label{Gmag_distribution}  
\end{figure}

{Figure \ref{Gmag_distribution} shows the distribution of the $G$-band magnitude for those 9273 WDs with a confirmed spectral class in the MWDD and for our 57,064 DNC candidates. As can be seen, the apparent magnitude is more concentrated in the second case, something expected since the MWDD WDs were observed using different facilities with a wider range of telescope sizes and limiting magnitudes. In any case, the $G$-band magnitude is not very different in DNC candidates (mainly contained within $\left[16, 20\right]$ mag) from the MWDD confirmed ones, so the DNC candidate classified spectra is not expected to be noisier. This is a necessary condition for good SOM performance. Otherwise, the morphological features produced by the noise can be interpreted as real spectral features by the SOM and these spectra could be incorrectly classified.}

\begin{figure*}
\gridline{\fig{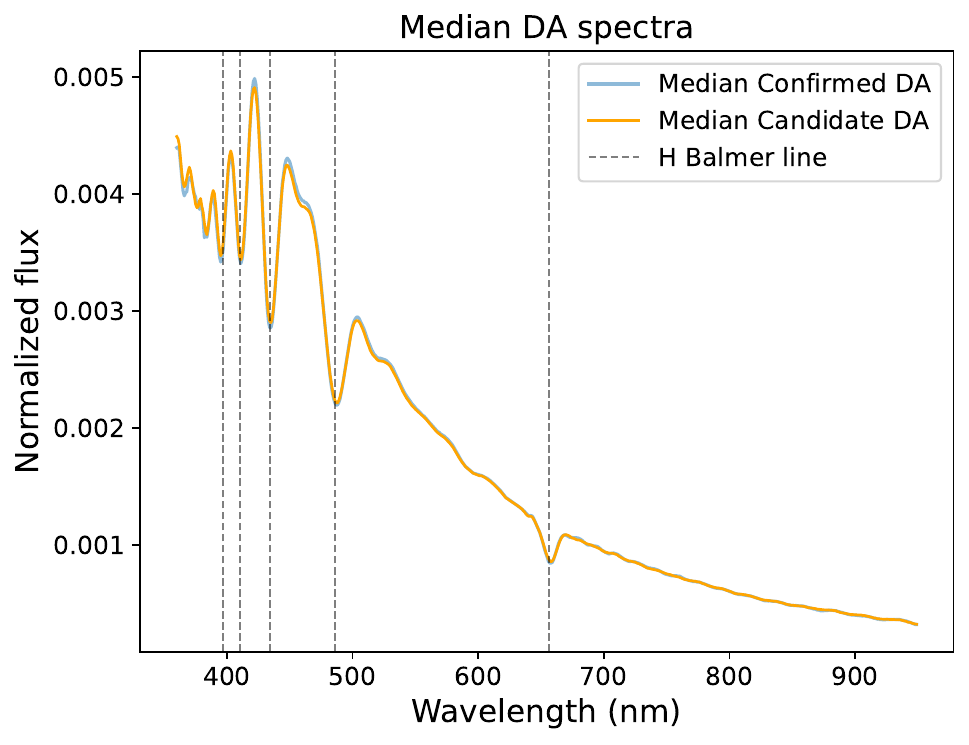}{0.495\textwidth}{(a) $z_{0,4}$ neuron normalized median spectra.} \label{DA_median_spectra}
          \fig{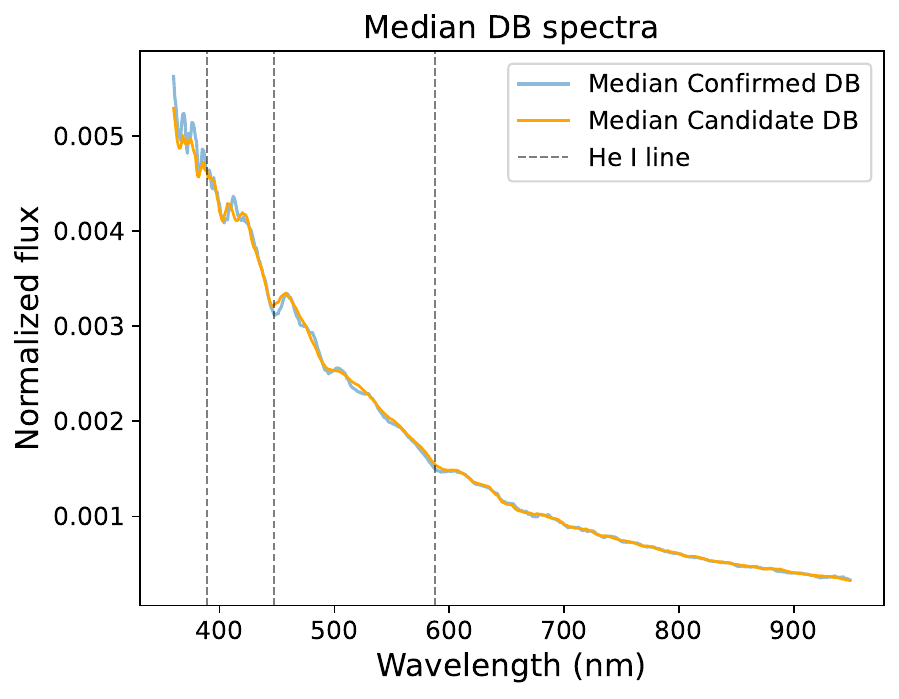}{0.495\textwidth}{(b) $z_{0,6}$ neuron normalized median spectra.} \label{DB_median_spectra}
}
\gridline{\fig{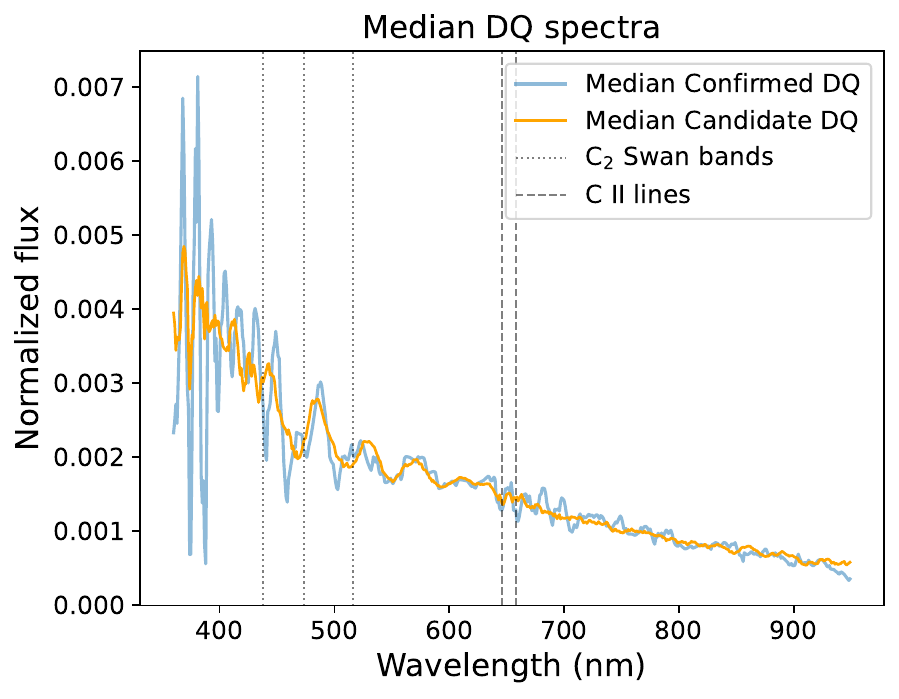}{0.495\textwidth}{(c) $z_{6,1}$ neuron normalized median spectra.} \label{DQ_median_spectra}
          \fig{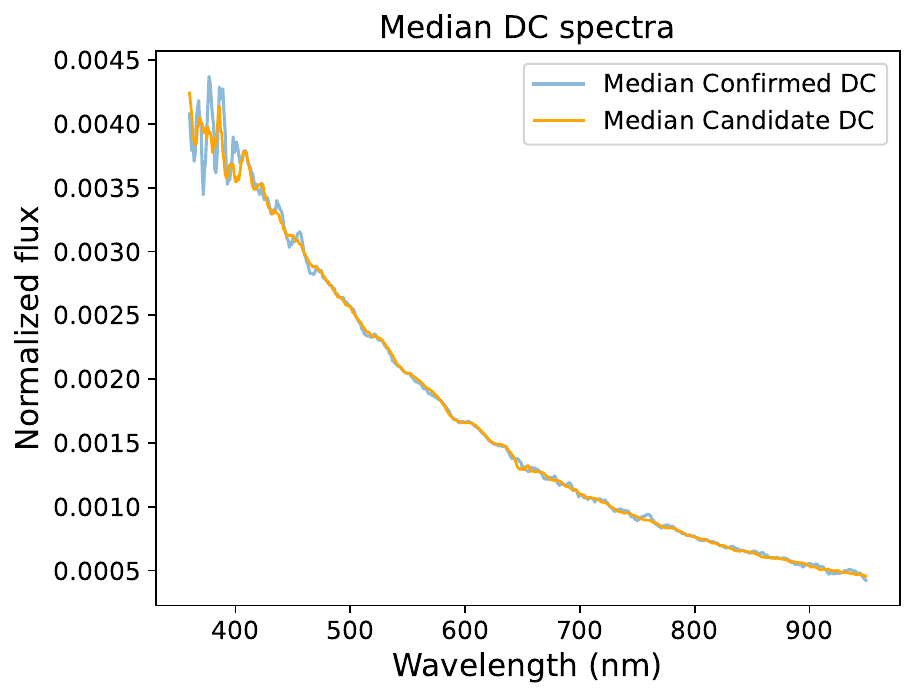}{0.495\textwidth}{(d) $z_{2,6}$ neuron normalized median spectra.}
}

\caption{\label{median_spectra} Normalized median spectra of WDs with confirmed spectral classification in the MWDD and identified candidates for several neurons.}
\end{figure*}

To visualize the reliability of the classification, {we have plotted the median externally calibrated spectrum of several spectral classes} using the \texttt{GaiaXPy}\footnote{\url{https://www.cosmos.esa.int/web/gaia/gaiaxpy}} Python library \citep{ruzmieres2024}. In Figure \ref{median_spectra}a, a comparison between the normalized median spectra of 972 MWDD confirmed DAs and that of 4379 DA candidates in neuron $z_{0,4}$ are shown overlapped. The H Balmer lines ($\lambda 397.0$, $\lambda 410.2$, $\lambda 434.0$, $\lambda 486.1$, $\lambda 656.3${; units in nanometers}) are plotted as dashed vertical lines. The same is done for the 253 confirmed DBs and 1478 DB candidates in neuron $z_{0,6}$ (Figure \ref{median_spectra}b) with their expected He I lines ($\lambda 388.9$, $\lambda 447.1$, $\lambda 587.6$). For DQ class, we used the neuron $z_{6,1}$ that, since it only contains 9 confirmed DQ sources with respect to 81 DQ candidates, looks noisier, as can be seen in Figure \ref{median_spectra}c where some Carbon lines for guidance are included. {For DC class, we plotted the median spectra of the neuron $z_{2,6}$ that, as can be seen in Figure \ref{median_spectra}d, is mainly featureless.}

\begin{figure*}
\gridline{\fig{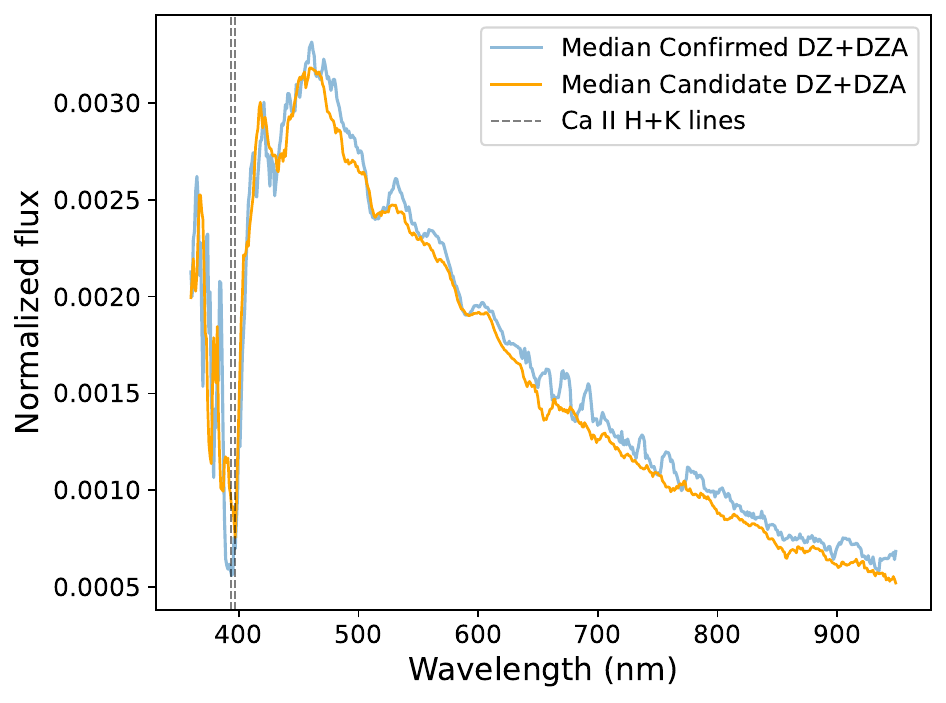}{0.495\textwidth}{(a) $z_{3,2}$ (DZA neuron) and $z_{7,2}$ (DZ neuron) combined and normalized median spectra.}
          \fig{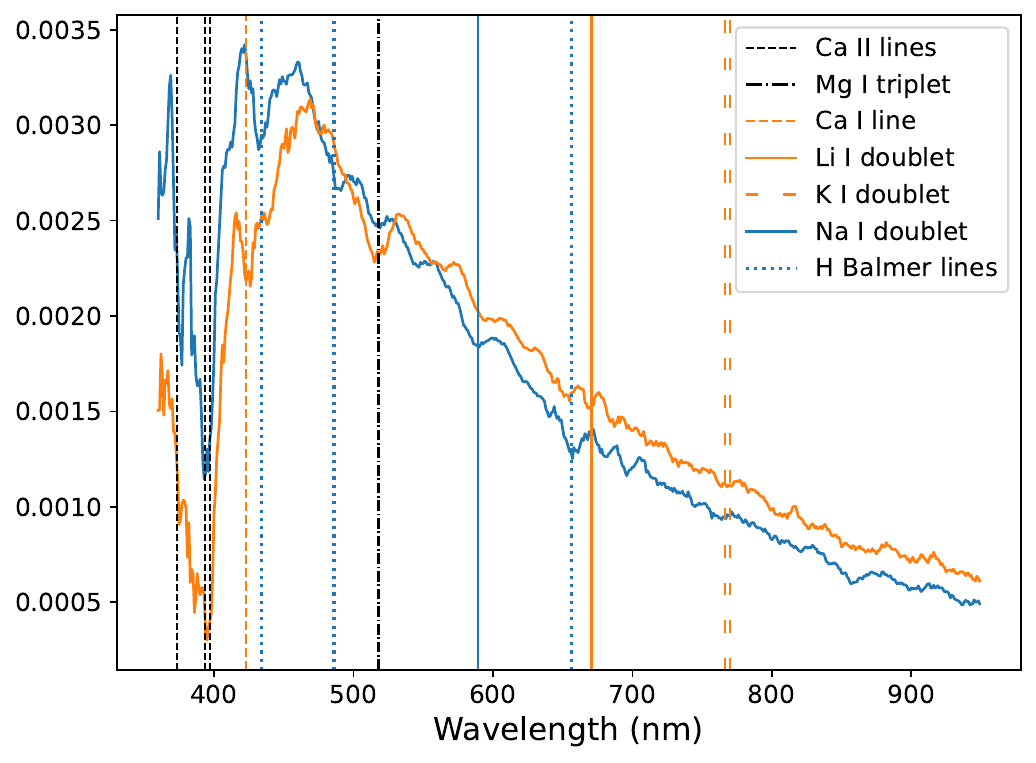}{0.495\textwidth}{(b)  DZA neuron (blue) and DZ neuron (orange) normalized median spectra. Spectral lines found in both neurons are in black, while those that are specific to neurons DZA and DZ are in blue and orange, respectively.}
}
\caption{\label{DZ_median_spectra} Normalized median spectra for the neurons {with identified polluted WDs.}}
\end{figure*}

{Regarding the polluted WDs, {they are mainly concentrated in neurons $z_{3,2}$ (hereafter, DZA neuron, with 249 sources) and $z_{7,2}$ (hereafter, DZ neuron, with 218 sources).} In Figure \ref{DZ_median_spectra}a the normalized median spectra of 68 MWDD confirmed DZs is shown overlapped to the normalized median spectra of the 399 DZ candidates. Furthermore, the desired Ca II H+K doublet ($\lambda 393.4$, $\lambda 397.9$) is clearly seen in both.} 

However, while both neurons are mainly populated {by WDs labeled in the MWDD as DZs ($\sim 80\%$ and $\sim 90\%$, calculated with respect to the total number of WDs with MWDD classification, respectively) the first one is mixed with $\sim 20 \%$ of MWDD DAs and DCs in equal parts, and the second one with $\sim 10\%$ of DCs.} {These indicate that other features than Calcium} may be present in these Gaia XP spectra. In Figure \ref{DZ_median_spectra}b the normalized median spectra of both neurons is plotted independently (DZA neuron in blue, and DZ neuron in orange) with several absorption lines for guidance. {Lines in common are plotted in black, those only found in the DZA neuron in blue, and those specific to the DZ neuron in orange.}

Besides the Ca II H\&K doublet, the  median spectra (Figure \ref{DZ_median_spectra}b, in blue), shows a plenty of H Balmer lines ($\lambda 434.0$, $\lambda 486.1$, and $\lambda 656.3$) indicating a population of H-rich polluted WDs in that neuron (hereafter, DZA neuron). An extra Ca II ($\lambda 373.7$) dash-dotted line and two more metals are present in the spectra, according to the features in the $\lambda 517.3-518.3$ Mg I triplet and in the $\lambda 589.0-589.6$ Na I doublet. 
Regarding the $z_{7,2}$ median spectra, we can see in Figure \ref{DZ_median_spectra}b (in orange) how the Balmer series disappeared, at the same time that other metals came to light, as suggested by the presence of the Ca I line ($\lambda 423.0$) and possible detections of Li I and K I doublets around $\lambda 670.8$ and $\lambda 766.5-769.9$, respectively. The absence of H and He I lines indicates that this neuron (hereafter, DZ neuron) may be populated by cool DZs.

\begin{figure}
\centering
    \includegraphics[width=0.495\textwidth]{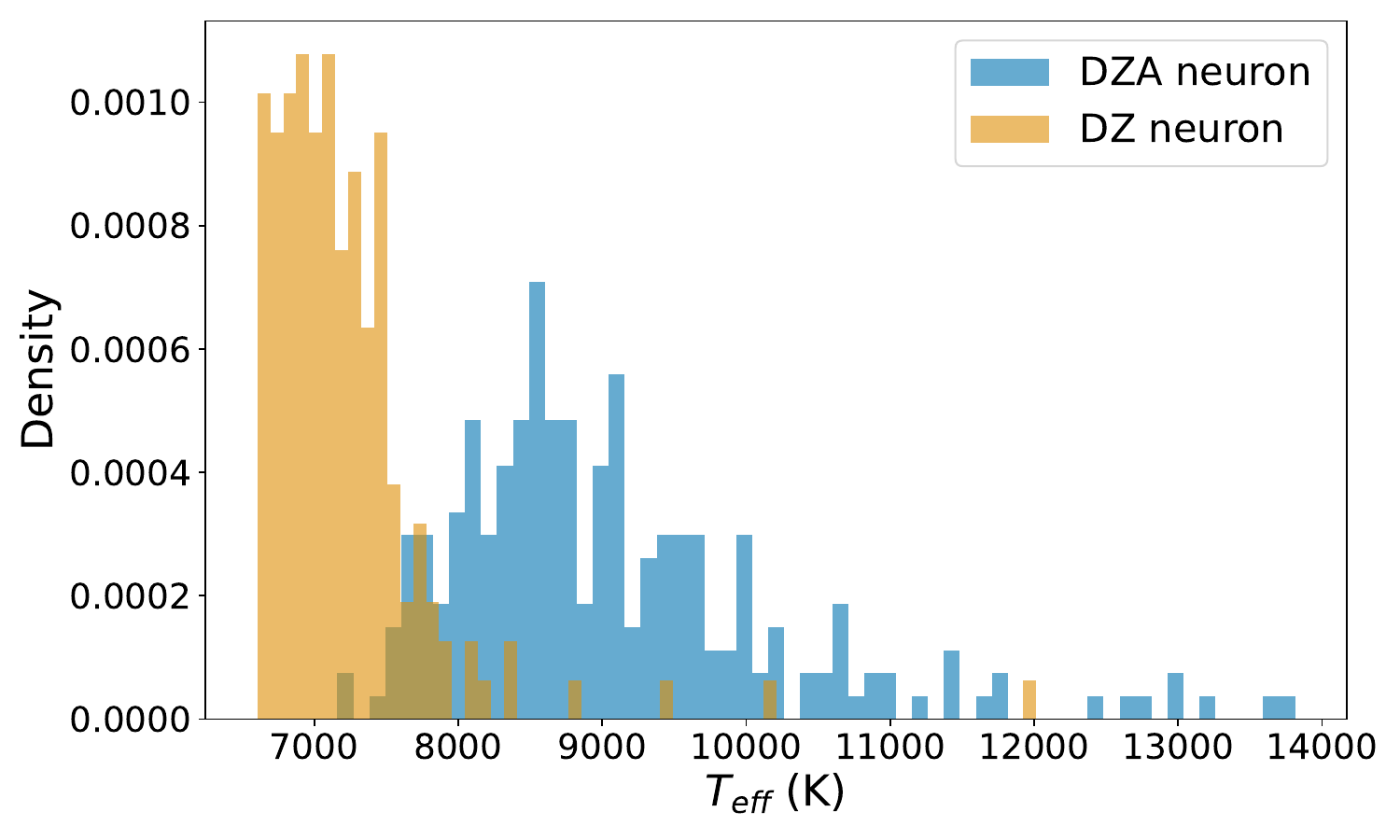}
    \caption{Density histograms of $T_{eff}$ assuming mixed H and He atmospheres from \citet{gentilefusillo2021} for DZA and DZ neurons in bins of 100 K.}
    \label{teff_DZ_distribution}    
\end{figure}

Indeed, the $T_{eff}$ distribution (obtained from \citet{gentilefusillo2021}, by assuming mixed H and He atmospheres), is greater in the DZA neuron, ranging from $7200$ K to $28,600$ K, with a mean of 9200 K. On the other side, the $T_{eff}$ distribution in the DZ neuron ranges from 6600 to $12,000$ K, with a mean of $7200$ K, which explains the absence of He I lines since below $11,000$ K there is not enough temperature to excite them, making the $z_{7,2}$ a cool DZ neuron. Metals displayed in the atmospheres of cool polluted WDs had to be accreted, as well as in almost all of WDs in the DZA neuron, since with a temperature $< 25,000$ K the radiative levitation could not be the cause \citep{chayeretal1995}. In Figure \ref{teff_DZ_distribution} we show the $T_{eff}$ density distribution for each neuron, {with the DZA neuron values up to the 99$^{th}$ percentile (corresponding to $T_{eff} < 14,400K $) to enhance visualization.}

\subsection{Comparison with other methodologies}\label{subsec:comparisonwith}

We have also compared our classification with supervised learning methods such as that of the \citet{garciazamoraetal2023} and \citet{vincentetal2024}.  \citet{garciazamoraetal2023} trained a Random Forest algorithm with MWDD labels to classify the WD 100 pc volume-limited sample of \citet{jimenezestebanetal2023}, while \citet{vincentetal2024} used an ensemble algorithm of decision trees, to classify the \citet{vincentetal2023} sample. 

\begin{figure*}
\gridline{\fig{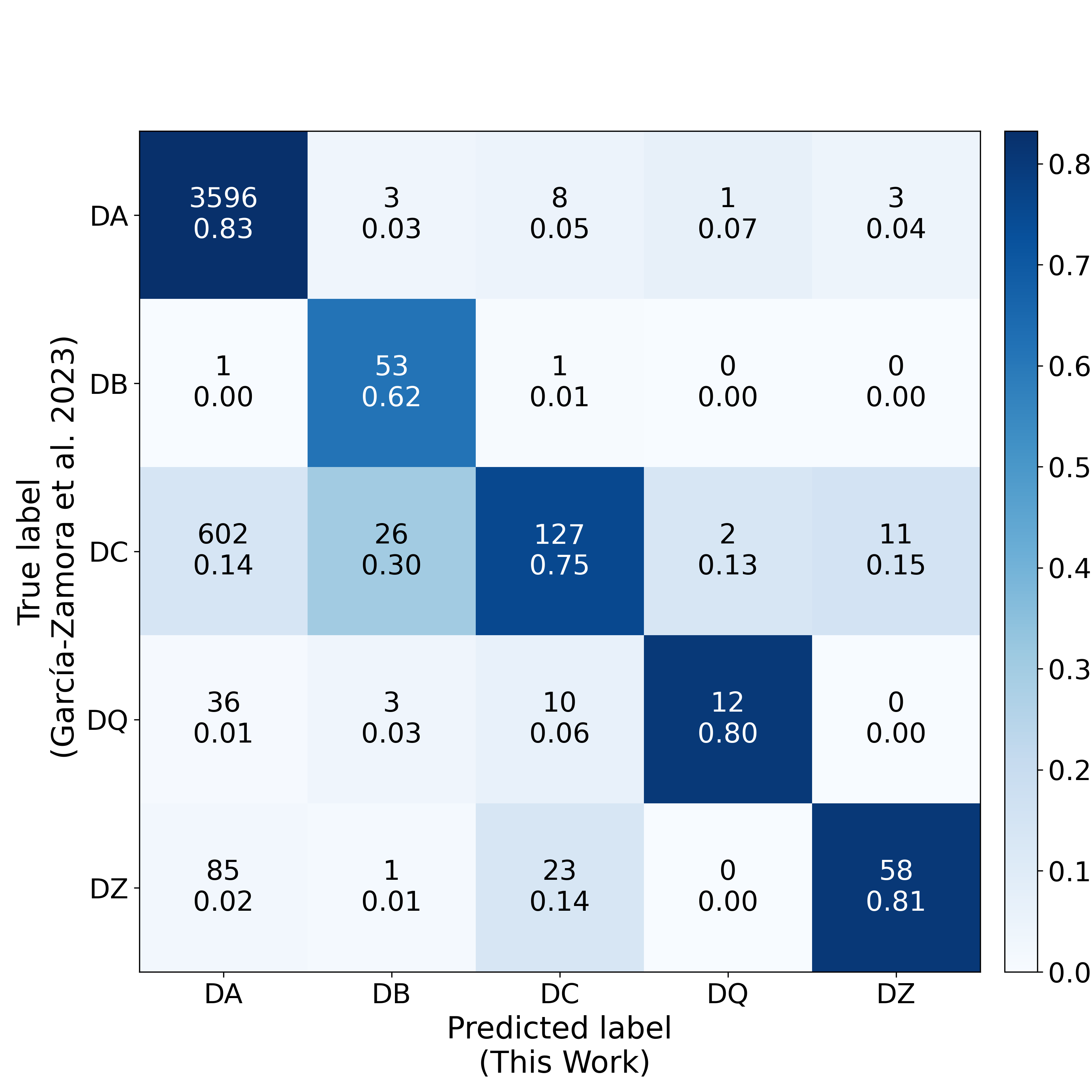}{0.495\textwidth}{(a)}\label{fig:garciazamoraSOM}
          \fig{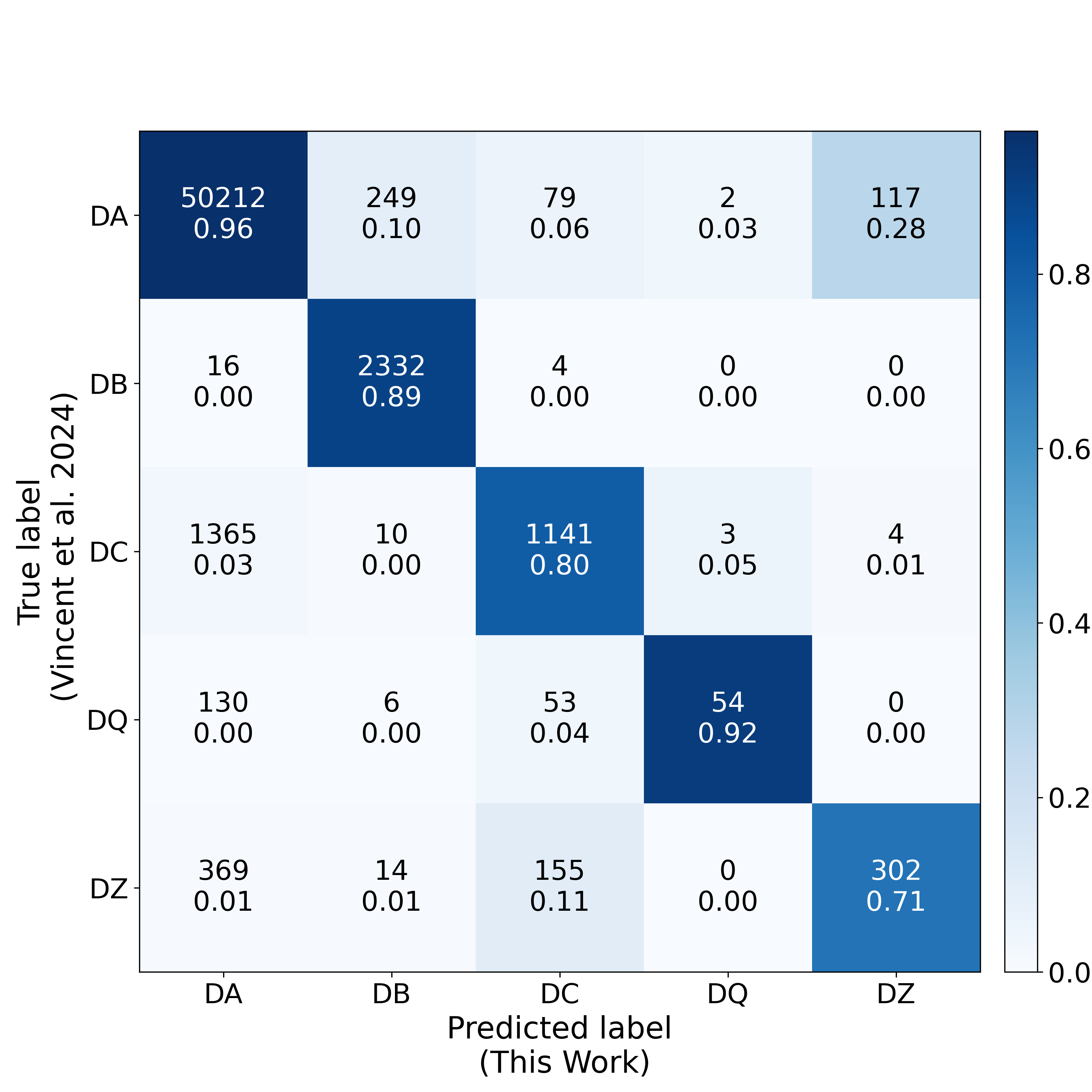}{0.495\textwidth}{(b)}\label{fig:vincentSOM}
}
\caption{\label{two_confusion_matrices} Confusion matrices of our work (the predicted labels) with \citet{garciazamoraetal2023} (a) and \citet{vincentetal2024} (b) (the true labels). The numbers in each cell have the same meaning as those in Figure \ref{fig:som_confusionmatrix}.}
\end{figure*}

To compare the results of our method with these supervised classifications, we cross-matched their catalogs with ours. {4662 WDs were found to be in common between \citet{garciazamoraetal2023} and our SOM. Since some of the candidates have secondary class classification in their catalog, we considered only their primary classification (so DAB, DAH, DAZ, \dots \: are all considered as DAs; DBAs, DBZs, \dots \: are all considered as DBs; and so on). On the other hand,} \citet{vincentetal2024} shows 56,617 candidates classified with reliable primary spectral types {i.e., not classified by the authors as uncertain (indicated by a ":" notation)} that are also in our catalog. The confusion matrices, with our labels considered as the predicted labels, are shown in Figures \ref{two_confusion_matrices}a and \ref{two_confusion_matrices}b, respectively.

{As can be seen, our results reveal a very good precision with respect to those of \citet{garciazamoraetal2023} except for our DBs, since a 30\% of them are classified in their work as DCs. On the other hand, the confusion matrix of our work with \citet{vincentetal2024} is nearly diagonal, showing also a good precision for all of the classes. Notwithstanding that, an excellent agreement requires an excellent recall, that is the main limitation of our SOM as stated before, and shown again here. 

{Indeed, we can see that, for instance, we identified a total of 72 DZs in the cross-match with \citet{garciazamoraetal2023}, of which} 58 are classified by them as DZs (81\%), 11 as DCs and 3 as DAs. However, of the 167 DZs in \citet{garciazamoraetal2023}, we only classify as such 58 of them (35\%), since more than a half are classified by the SOM as DAs. The same scenario occurs with \citet{vincentetal2024} cross-match, where of the 423 WD candidates identified as DZs by the SOM, 302 (71\%) are also classified as DZs in their work. However, of the 840 WDs labeled as DZs by them, only 302 are identified as DZs in our work (36\%).}

{This results are in high consonance with those shown in \S \ref{subsec:spectralclassification} (see Figure \ref{fig:som_confusionmatrix} and Table \ref{tab:som_metrics}) and reinforce the fact that the main strength of the SOM is its high precision, but at the cost of having a very low recall.}

{Of the 467 sources classified as polluted, 136 were found in the Montreal White Dwarf Database but only 68 have been classified as polluted WDs up to the 3rd of July 2024.}
{Of the remaining 399 DZ candidates, 37 ($9\%$) were classified as DZ in \citet{garciazamoraetal2023}, and 236 ($59\%$) in \citet{vincentetal2024} (36 of them are also in common with \citealt{garciazamoraetal2023}).}

{We have also cross-matched our sample with the polluted WD catalog of \citet{badenasagustietal2024} and two only sources matched with our sample, they are also in \citet{garciazamoraetal2023} and \citet{vincentetal2024}.}

During the preparation of this manuscript, \citet{kaoetal2024} used UMAP to perform an unsupervised classification of the GF+21 catalog, obtaining a sample of 375 polluted candidates. {Although they did not publish that sample in their paper, they provide the UMAP projected 2D coordinates for each source. Thus, we were able to cross-match their catalogue with our sample.}

\begin{figure}[h]
    \centering
    \includegraphics[width=0.49\textwidth]{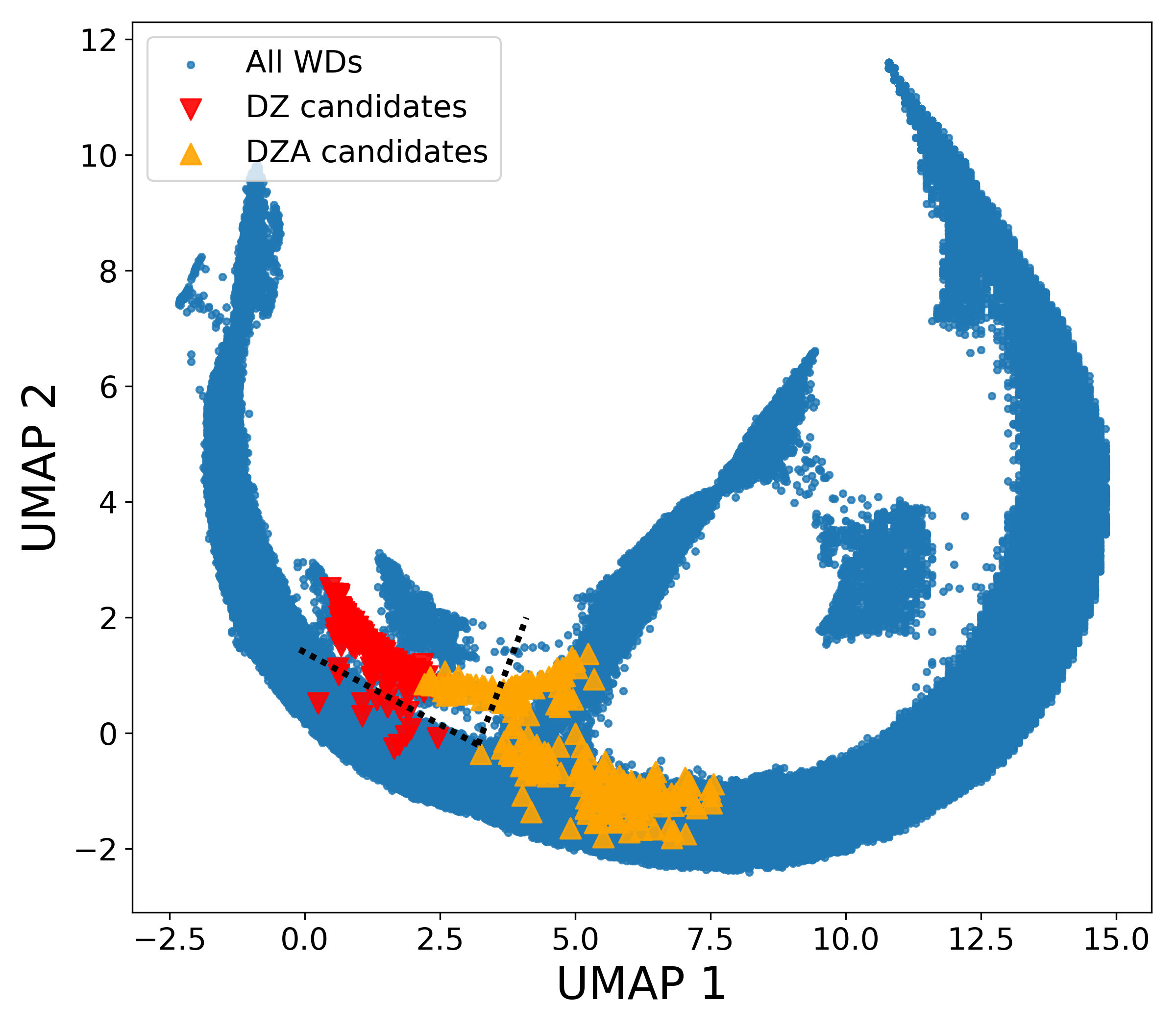}
    \caption{UMAP of \citet{kaoetal2024} with all of their WD candidates in blue and our DZ and DZA candidates in red and orange, respectively. The black dotted lines served us to discriminate the polluted WD candidates identified in their work from those identified here.}
    \label{UMAP_SOM_xmatch}
\end{figure}

{Of our 467 DZ candidates, 456 of them have positions in their UMAP, and nine of the remaining 11 are not in the previous catalogs. In Figure \ref{UMAP_SOM_xmatch} we show the UMAP of \citet{kaoetal2024} in blue and our DZ and DZA candidates in red and orange, respectively. As can be seen in Figure 4 of their work, the polluted WDs they have found are concentrated in a DZ island in the UMAP. Our DZ and DZA candidates are distributed through that island but also in other parts of the UMAP so they are not in the \citet{kaoetal2024} DZ sample.}

{To roughly determine how many of our candidates are actually new WD candidates, we have set the following conditions \eqref{lineA} and \eqref{lineB} for the UMAP coordinates (\texttt{UMAP1} and \texttt{UMAP2}) so that we filtered the candidates of our sample that are below the two black dotted lines depicted in Figure \ref{UMAP_SOM_xmatch} and that wrap around the DZ island:}
{\begin{equation} 
\label{lineA}
\texttt{UMAP2} < 2.44 \times \texttt{UMAP1} - 8.02, \text{ if }
\texttt{UMAP1} \geq 3.20
\end{equation}
\begin{equation} \label{lineB}
\texttt{UMAP2} < -0.50 \times \texttt{UMAP1} + 1.40, \text{ if }
\texttt{UMAP1} < 3.20
\end{equation}}
{We find that from the 456 of our polluted WD candidates that are in the UMAP, 225 sources (13 DZ candidates and 212 DZA candidates) are out of the \citet{kaoetal2024} DZ island. Moreover, the candidates in the {DZ neuron have more metal species identified in their Gaia spectra than those} of the DZA neuron, which explains that we have more of them in common with their work, since they also found up to five different metal species.}

{This together with all of the cross-matches that we have described before, gives us confidence to report that at least 143 of our polluted WD candidates (133 DZAs and 10 DZs) {are completely new discoveries}.}


{As mentioned before, (see Table \ref{tab:som_metrics} as well), the main limitation of the SOMs presented in this work is their lack of sensitivity or recall, mainly because of the labeling procedure. {Despite of that,} we demonstrated here that both the SOM filter as well as the classification SOM were able {to recover many high-confidence polluted WDs} that previous supervised and unsupervised learning works overlooked. Indeed, the combination of dimensionality reduction and clustering skills of the SOM allowed to identify a population of new DZA candidates that the UMAP missed.}

{Another important advantage of the clustering characteristic in the SOM is that} it allows us to perform statistics within each neuron in order to obtain a golden sample with only the best candidates for each spectral class. 

For instance, if we compute the intra-neuron distance (IND) distribution for each neuron as the distance of each spectrum to the mean spectrum of that neuron, we can obtain an internal validation of our procedure: a neuron is a reliable cluster if and only if the INDs are low, showing an assymetric left-skewed Gaussian shape. Appendix 1 displays, in Figure \ref{fig:filter_SOM_IND}, the distribution of distances achieved by mapping our sample of 104,844 sources to the SOM map pretrained with SDSS sources in Figure \ref{fig:gaiasdsss_som}, while in Figure \ref{fig:classification_SOM_IND} we show the IND distribution for each neuron in the map presented in Figure \ref{fig:som_candidates}. The distance distribution mirrors that of the original maps, demonstrating the method's effectiveness and enabling reliable classification. Moreover, we can choose the most reliable objects as those that have the lowest INDs within their own neuron.

This work outlines the methodology used to identify a set of bona fide polluted WD candidates, detailing the steps taken to ensure the work's reproducibility. The list of candidate objects will be validated with high-resolution observations.

\section{Conclusions}\label{sec:conclusions}

We have demonstrated in this work the power of unsupervised learning to filter data by obtaining a clean sample of $66,337$ WD candidates from the Gaia white dwarf catalog of \citet{gentilefusillo2021} using alternative input data (Gaia XP spectra instead of the Gaia CMD) and methodology (unsupervised machine learning with SOM). This has led us to a less complete but more robust and high-confidence sample of WDs, as the XP coefficients store more information than just $G$, $G_{BP}$, $G_{RP}$ magnitudes. By applying the SOM algorithm to the GF+21 sample of WD candidates with Gaia XP spectra as input data, we can exclude contaminants in a more robust and reliable way than by using the $P_{WD}$, which is solely based on the position of the candidate in the H-R diagram, as well as classify the true WDs according to their spectral morphology. Moreover, the fact that 791 sources of our sample appear as low-confidence ($P_{WD} < 0.75$) WDs in the GF+21 catalog proves our methodology to be a useful tool for identifying hidden but reliable WDs.

Regarding the spectral classification, unsupervised Self-Organizing Maps have shown a similar performance that in recent supervised learning works, with high precision for DA, DB, DQ, and DZ white dwarfs. This strongly justifies the use of Self-Organizing Maps as they provide a natural and useful way to group similar spectra and, at the same time, to label new data, by performing statistics within each neuron.

{This method allowed us to identify, with high confidence, 143 new polluted WD candidates that show spectral features of  several metals (namely, Ca, Mg, Na, Li, and K), and even to distinguish between DZ and DZA subtypes.} In order to confirm those candidates and to delve into other interesting neurons (such as DQ and DXZ subtypes) follow-up spectroscopy of the best candidates will be performed in the near future.

\begin{acknowledgements}
{We warmly thank the anonymous referee whose insightful comments have greatly improved this paper.} This work has made use of data from the European Space Agency (ESA) Gaia mission and processed by the Gaia Data Processing and Analysis Consortium (DPAC). Funding for the DPAC has been provided by national institutions, in particular the institutions participating in the Gaia Multilateral Agreement. {This work has made use of the Python package GaiaXPy, developed and maintained by members of the Gaia Data Processing and Analysis Consortium (DPAC) and in  particular, Coordination Unit 5 (CU5), and the Data Processing Centre located at the Institute of Astronomy, Cambridge, UK (DPCI).} This research was funded by the Horizon Europe [HORIZON-CL4-2023-SPACE-01-71] SPACIOUS project, Grant Agreement no. 101135205, the Spanish Ministry of Science MCIN / AEI / 10.13039 / 501100011033, and the European Union FEDER through the coordinated grant PID2021-122842OB-C22. We also acknowledge support from the Xunta de Galicia and the European Union (FEDER Galicia 2021-2027 Program) Ref. ED431B 2024/21, ED431B 2024/02, and CITIC ED431G 2023/01. X.P. acknowledges financial support from the Spanish National Programme for the Promotion of Talent and its Employability grant PRE2022-104959 cofunded by the European Social Fund. Funding from Spanish Ministry project  PID2021-127289NB-100 is also acknowledged.
\end{acknowledgements}


\begin{thebibliography}{99}

\begin{small}

\bibitem[\'Alvarez et al.(2022)]{alvarezetal2022} \'Alvarez, M.A., Dafonte, C., Manteiga, M. et al. 2022, Neural Comput \& Applic, 34, 1993–2006

\bibitem[Andrae et al.(2023)]{andraeetal2023} Andrae, R., Fouesneau, M., Sordo, R., et al. 2023, A\&A, 674, id.A27, 22.

{\bibitem[Badenas-Agusti et al.(2024)]{badenasagustietal2024} Badenas-Agusti, M., Vanderburg, A., Blouin, S., et al.\ 2024, \mnras, 527, 4515. doi:10.1093/mnras/stad3362}

\bibitem[Carvalho et al.(2016)]{carvalhoetal2016} Carvalho, A., Marinho Jr., R. M., Malheiro, M., et al 2016, J. Phys.: Conf. Ser. 706 052016

\bibitem[Carrasco and Brunner(2014))]{carrascoandbrunner2014} Carrasco, K. M., Brunner, R. J. 2014, MNRAS, 438(4), 3409–3421

\bibitem[Carrasco et al.(2021)]{carrascoetal2021} Carrasco, J. M., Weiler, M., Jordi, C., et al. 2021, A\&A, 652, A86

\bibitem[Chayer et al.(1995)]{chayeretal1995} Chayer, P., Fontaine, G., \& Wesemael, F.\ 1995, \apjs, 99, 189. doi:10.1086/192184

\bibitem[Dafonte et al.(2018)]{dafonteetal2018} Dafonte, C., Garabato, D., Álvarez, M.A., Manteiga, M. 2018, Sensors, 18, 1419

\bibitem[De Angeli et al.(2023)]{deangelietal2023} De Angeli, F., Weiler, M., Montegriffo, P., et al. 2023, A\&A, 674, A2

\bibitem[Delchambre et al.(2023)]{delchambreetal2023} Delchambre, L., Bailer-Jones, C.~A.~L., Bellas-Velidis, I., et al.\ 2023, \aap, 674, A31. doi:10.1051/0004-6361/202243423


\bibitem[Dufour et al.(2016)]{dufouretal2016} Dufour, P., Blouin, S., et al. 2016, arXiv:1610.00986 [astro-ph.SR]

\bibitem[Echeverry et al.(2022)]{echeverryetal2022} Echeverry, D., Torres, S., Rebassa-Mansergas, A., et al. 2022, A\&A, 667, A144

\bibitem[Farihi et al.(2010)]{farihietal2010} Farihi, J., Barstow, M. A., Redfield, S., et al. 2010, MNRAS, 404, 2123

\bibitem[Fustes et al.(2013a)]{fustesetal2013a} Fustes, D., Dafonte, C., Arcay, B. et al. 2013, Expert Syst Appl, 40(5), 1530–1541.

\bibitem[Fustes et al.(2013b)]{fustesetal2013b} 
Fustes, D., Manteiga, M., Dafonte, C. et al. 2013, A\&A, A7, 10.

\bibitem[Gaia Collaboration(2023)]{gaiacollaboration2023} Vallenari, A., Brown, A.G.A., Prusti, T., et al. 2023, A\&A 674, A1

\bibitem[Garabato(2020)]{garabato2020} Garabato, D. 2020, PhD thesis. University of A Coruña

\bibitem[García-Zamora et al.(2023)]{garciazamoraetal2023}
Garcia-Zamora   E. M., Torres   S., Rebassa-Mansergas   A., 2023, A\&A, 679, A127 

\bibitem[Geach(2012)]{geach2012} Geach,. J. E. 2012, MNRAS, 419, 2633–2645

\bibitem[Gentile-Fusillo et al.(2019)]{gentilefusillo2019}
Gentile Fusillo, N.~P., Tremblay, P.-E., G{\"a}nsicke, B.~T., et al.\ 2019, \mnras, 482, 4570. doi:10.1093/mnras/sty3016
{
\bibitem[Gentile-Fusillo et al.(2021)]{gentilefusillo2021}
Gentile Fusillo, N.~P., Tremblay, P.-E., Cukanovaite, E., et al.\ 2021, \mnras, 508, 3877. doi:10.1093/mnras/stab2672
}
\bibitem[Ginsburg et al.(2019)]{ginsburgetal2019} Ginsburg, A., Sip{\H{o}}cz, B.~M., Brasseur, C.~E., et al.\ 2019, \aj, 157, 98. doi:10.3847/1538-3881/aafc33

\bibitem[Golovin et al.(2024)]{golovinetal2024} Golovin, A., Reffert, S., Vani, A., et al.\ 2024, \aap, 683, A33. doi:10.1051/0004-6361/202347767


\bibitem[Iben et al.(1997)]{ibenetal1997} Iben I. J., Ritossa C., Garcia-Berro E., 1997, ApJ, 489, 772

\bibitem[Izquierdo et al.(2020)]{izquierdoetal2020} Izquierdo, P., Toloza, O, Gänsicke, B. T., et al. 2020, MNRAS, 501(3), 4276-4288

\bibitem[Jim\'enez-Esteban et al.(2023)]{jimenezestebanetal2023} Jiménez-Esteban, F. M., Torres, S., Rebassa-Mansergas, A., et al. 2023, MNRAS, 518, 5106.
{
\bibitem[Kao et al.(2024)]{kaoetal2024} Kao, M.~L., Hawkins, K., Rogers, L.~K., et al.\ 2024, \apj, 970, 181. doi:10.3847/1538-4357/ad5d6e
}

\bibitem[Klein et al.(2021)]{kleinetal2021} Klein, B. L., Doyle, A. E., Zuckerman, B., et al. 2021, ApJ, 914(1), id.61, 17

\bibitem[Koester \& Wilken(2006)]{koesteretal2006} Koester, D., Wilken, D. 2006, A\&A, 453, 1051.

\bibitem[Koester(2009)]{koester2009} Koester, D. 2009, A\&A, 498(2), 517-525.

\bibitem[Kohonen(1982)]{kohonen1982}  Kohonen, T. 1982, Biol. Cybern. 43, 59-69

\bibitem[Lema{{\^i}}tre et al.(2017)]{lemaitre2017} Lema{{\^i}}tre, G., Nogueira, F., Aridas, C. K. 2017, JMLR, 18(17), 1-5.

\bibitem[Lindegren et al.(2018)]{lindegrenetal2018} Lindegren, L., Hern\'andez, J., Bombrun, A., et al. 2018. A\&A, 616, id.A2, 25

\bibitem[Maldonado et al.(2020)]{maldonadoetal2020} Maldonado, R. F., Villaver, E., Mustill, A. J., et al. MNRAS, 499(2), 1854-1869

\bibitem[Maldonado et al.(2021)]{maldonadoetal2021} Maldonado, R. F., Villaver, E., Mustill, A. J., et al. 2021. MNRAS, 501(1), L43-L48

\bibitem[Mustill et al.(2018)]{mustilletal2018} Mustill, A. J., Villaver, E., Veras, D., et al. 2018, MNRAS, 476(3), 3939-3955

\bibitem[Naim et al.(2009)]{naimetal2009} Naim, A., Ratnatunga, U., Griffiths, E. 2009, ApJ Suppl
Series 111, 357

\bibitem[Ordoñez-Blanco et al.(2010)]{ordonezetal2010} Ordoñe-Blanco, D., Arcay, B., Dafonte, C. et al. 2010,
Lect Notes Essays Astrophys, 4, 97–102

\bibitem[Pallas-Quintela et al.(2023)]{pallasquintela2023} Pallas-Quintela, L., Garabato, D., Manteiga, M., Dafonte, C. 2023, Highlights of Spanish Astrophysics XI, Proceedings of the XV Scientific Meeting of the Spanish Astronomical Society held on September 4--9, 2022, in La Laguna, Spain. M. Manteiga, L. Bellot, P. Benavidez, A. de Lorenzo-Cáceres, M. A. Fuente, M. J. Martínez, M. Vázquez Acosta, C. Dafonte (eds.)

\bibitem[Pelletier et al.(1986)]{pelletieretal1986} Pelletier, C., Fontaine, G., Wesemael, F., Michaud, G., Wegner, G. 1986, ApJ, 307, 242

\bibitem[Riello et al.(2021)]{rielloetal2021} Riello et al. 2021, A\&A, 649, id.A3, 33 pp.
{\bibitem[Ruz-Mieres(2024)]{ruzmieres2024} Ruz-Mieres, D. 2024, gaia-dpci/GaiaXPy: GaiaXPy v2.1.2, doi: 10.5281/zenodo.11617977}

\bibitem[Sion et al.(1983)]{sionetal1983} Sion, E. M., Greenstein, J. L., Landstreet, J. D., et al. 1983, ApJ, 269, 253

\bibitem[Swan et al.(2023)]{swanetal2023} Swan, A., Farihi, J., Su, K. L-U., Desch, S. J. 2024, MNRAS: letters, 529(1), L41–L46

\bibitem[Torres et al.(1998)]{torresetal1998} Torres, S., García-Becerro, E., Isern, J. 1998, ApJ, 508, L71

\bibitem[Trierweiler et al.(2023)]{trierweileretal2023} Trierweiler, . L., Doyle, A. E., Young, E. D. 2023, PSJ, 4(8), id.136, 13

\bibitem[Veras et al.(2024)]{verasetal2024} Veras, D., Mustill, A., Bonsor, A. 2024,  
Reviews in Mineralogy and Geochemistry, 90(1), 141-170

\bibitem[Vettigli(2018)]{vettigli2018} Vettigli, G. 2018, \url{https://github.com/JustGlowing/minisom/}

\bibitem[Vincent et al.(2023)]{vincentetal2023} Vincent, O., Bergeron, P., Dufour, P. 2023, MNRAS, 521, 760

\bibitem[Vincent et al.(2024)]{vincentetal2024} Vincent, O., Barstow, M. A., Jordan, S., et al. 2024, A\&A, 682, A5

\bibitem[Way and Klose(2012)]{wayandklose2012} Way, M., Klose, C. 2012, PASP, 124

\bibitem[Weiler et al.(2023)]{weileretal2023} Weiler, M., Carrasco, J. M., Fabricius, C., Jordi, C. 2023, A\&A 671, A52

\bibitem[Xu et al.(2024)]{xuetal2024} Xu, S., Rogers, L., Blouin, S. 2024, arXiv:2404.15425

\bibitem[Zuckerman et al.(2007)]{zuckermanetal2007} Zuckerman, B., Koester, D., Melis, C., et al. 2007, ApJ, 671(1), 872-877

\end{small}

\end{thebibliography}

\bibliographystyle{apalike}

\appendix

\section{Intra-neuron distance distributions}

\begin{figure}[h] 
    \centering
    \includegraphics[width=\textwidth]{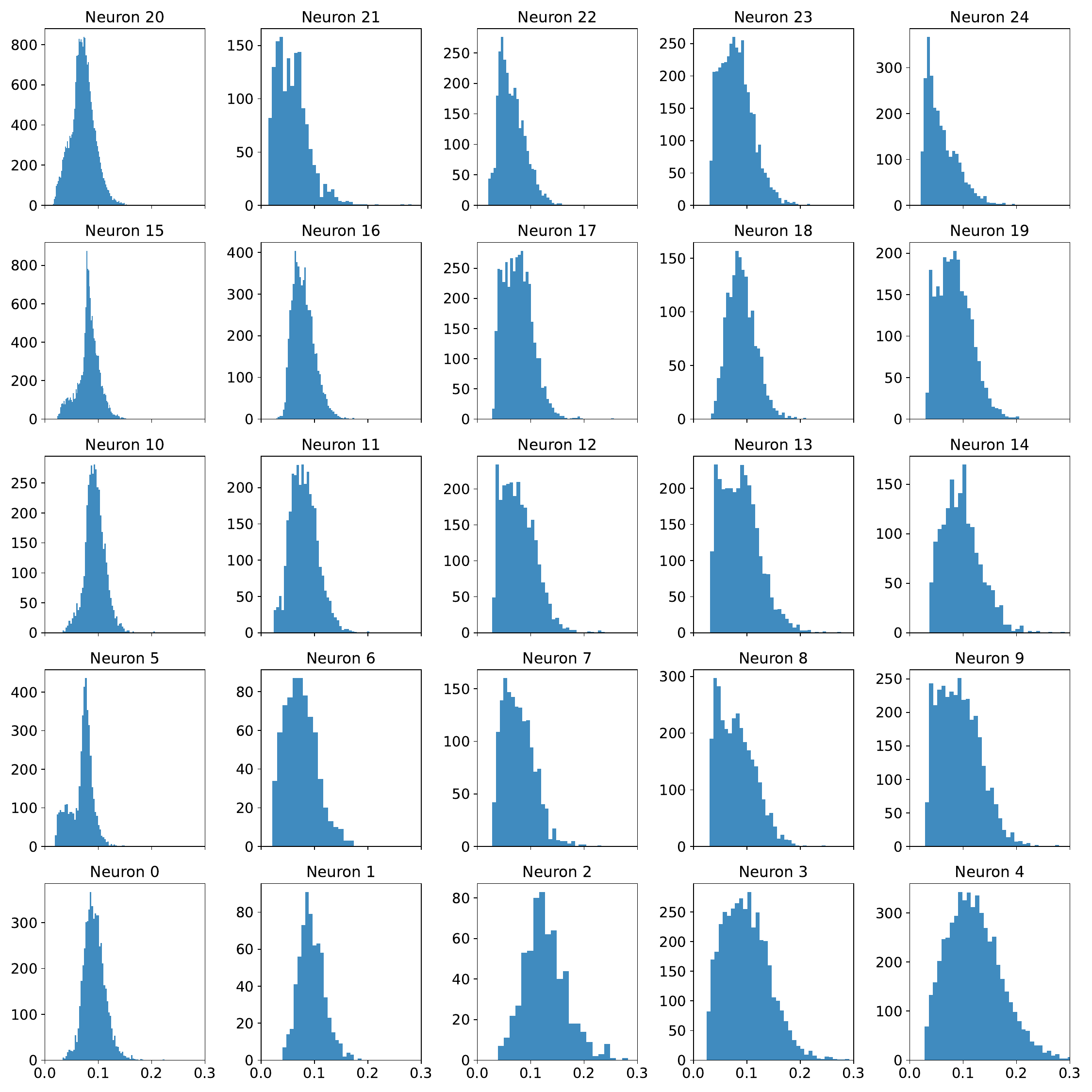}
    \caption{Intra-Neuron Distance distribution for the SOM used for filtering out contaminants from reliable WDs (see Figure \ref{fig:gaiasdsss_som}). Neuron 0 is $z_{0,0}$, 1 is $z_{0,1}$, {5} is $z_{1,0}$, and so on.}
    \label{fig:filter_SOM_IND}
\end{figure}

\begin{figure}[h]
    \centering
    \includegraphics[width=\textwidth]{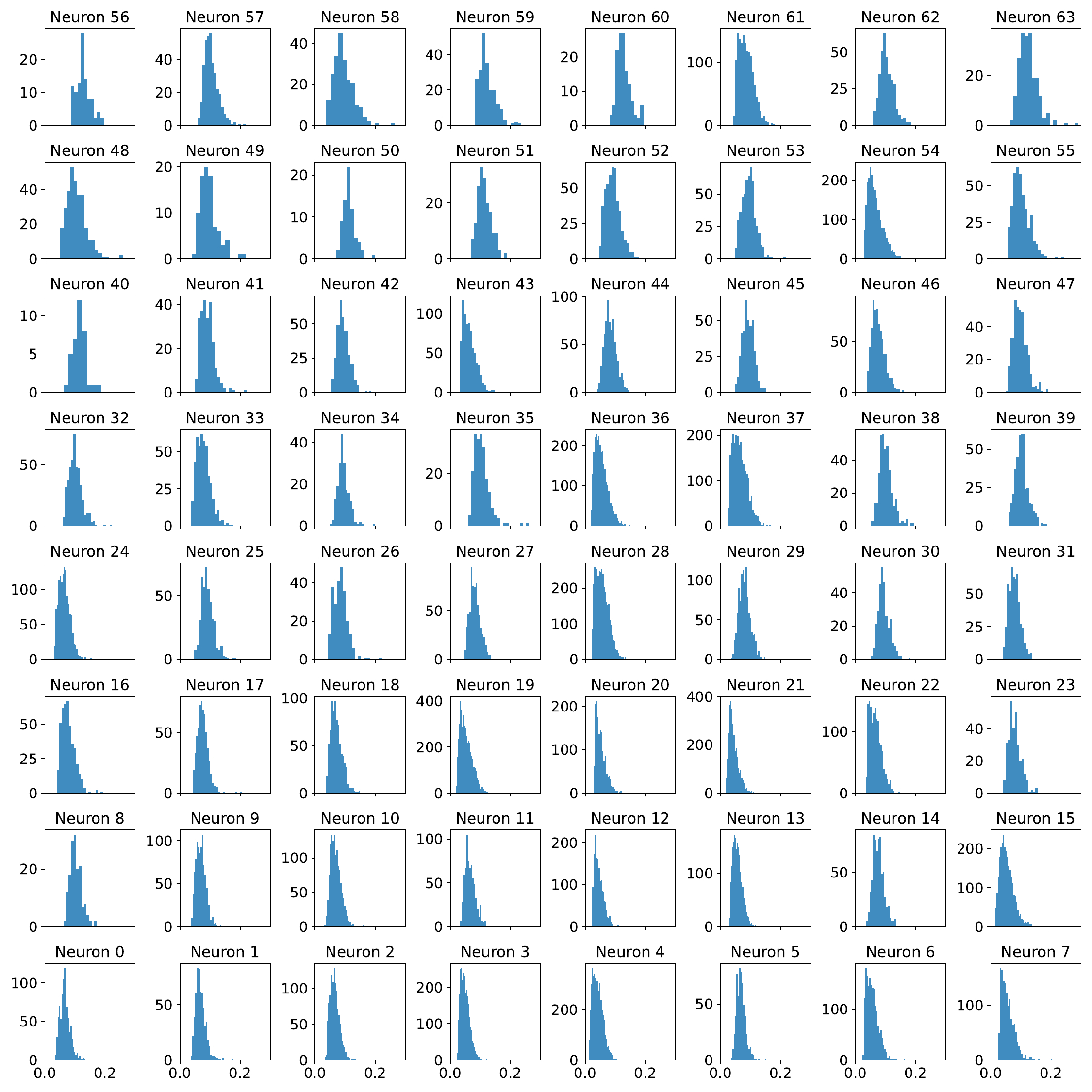}
    \caption{Intra-Neuron Distance distribution for the SOM used for spectral classification (see Figure \ref{fig:som_candidates}). Neuron 0 is $z_{0,0}$, 1 is $z_{0,1}$, 8 is $z_{1,0}$, and so on.}
    \label{fig:classification_SOM_IND}
\end{figure}

\end{document}